\documentclass[useAMS,usenatbib,usegraphicx]{mn2e}
\usepackage{amsmath}
\usepackage{pdflscape,longtable}
\usepackage{times}
\title[Properties of SNe Ia in rich clusters]{Properties of type Ia supernovae inside rich galaxy clusters}
\author[H. S. Xavier et al.]
{Henrique S. Xavier$^{1,2}$\thanks{E-mail: hsxavier@if.usp.br}, Ravi R. Gupta$^2$, Masao Sako$^2$, Chris B. D'Andrea$^{3}$, 
\newauthor Joshua A. Frieman$^{4,5,6}$, Lluis Galbany$^{7,8}$, Peter M. Garnavich$^9$, John Marriner$^6$, 
\newauthor Robert C. Nichol$^{3}$, Matthew D. Olmstead$^{10}$, Donald P. Schneider$^{11,12}$,
\newauthor Mathew Smith$^{13}$\\
$^{1}$ Instituto de Fisica, Universidade de Sao Paulo, Rua do Matao, Travessa R, 187, Sao Paulo, SP 05508-090, Brazil\\
$^{2}$ Department of Physics \& Astronomy, University of Pennsylvania, 209 South 33rd Street, Philadelphia, PA 19104, USA\\
$^{3}$ Institute of Cosmology and Gravitation, University of Portsmouth, Portsmouth, PO1 3FX, UK\\
$^{4}$ Department of Astronomy and Astrophysics, University of Chicago, 5640 South Ellis Avenue, Chicago, IL 60637, USA\\
$^{5}$ Kavli Institute for Cosmological Physics, University of Chicago, 5640 South Ellis Avenue Chicago, IL 60637, USA\\
$^{6}$ Center for Particle Astrophysics, Fermi National Accelerator Laboratory, P.O. Box 500, Batavia, IL 60510, USA\\
$^{7}$ Institut de Fisica d’Altes Energies, Universitat Autonoma de Barcelona, E-08193 Bellaterra (Barcelona), Spain\\
$^{8}$ Centro Multidisciplinar de Astrofisica, Instituto Superior Tecnico, Av. Rovisco Pais 1, 1049-001 Lisbon, Portugal\\
$^{9}$ Department of Physics, University of Notre Dame, 225 Nieuwland Science Hall, Notre Dame, IN 46556, USA\\ 
$^{10}$ Department of Physics and Astronomy, University of Utah, Salt Lake City, UT 84112, USA\\
$^{11}$ Department of Astronomy and Astrophysics, The Pennsylvania State University, University Park, PA 16802, USA\\
$^{12}$ Institute for Gravitation and the Cosmos, The Pennsylvania State University, University Park, PA 16802, USA\\
$^{13}$ Department of Physics, University of the Western Cape, Bellville, Cape Town, 7535, South Africa}

\begin{document}

\maketitle

\begin{abstract}

We used the GMBCG galaxy cluster catalogue and SDSS--II supernovae data with 
redshifts measured by the BOSS project to identify 48 SNe Ia residing in 
rich galaxy clusters and compare their properties with 1015 SNe Ia in the field. Their 
light curves were parametrised by the SALT2 model and the significance 
of the observed differences was assessed by a resampling technique. 
To test our samples and methods, we first looked for known differences 
between SNe Ia residing in active and passive galaxies. We confirm that 
passive galaxies host SNe Ia with smaller stretch, weaker colour--luminosity 
relation [$\beta$ of 2.54(22) against 3.35(14)], and that are $\sim 0.1$ mag more 
luminous after stretch and colour corrections. We show that only 0.02 per cent 
of random samples drawn from our set of SNe Ia in active galaxies can reach 
these values. Reported differences in the Hubble residuals scatter could not 
be detected, possibly due to the exclusion of outliers. We then show that, while 
most field and cluster SNe Ia properties are compatible at the current level, 
their stretch distributions are different ($\sim 3\sigma$): besides having a higher 
concentration of passive galaxies than the field, the cluster's passive galaxies 
host SNe Ia with an average stretch even smaller than those in field 
passive galaxies (at 95 per cent confidence). 
We argue that the older age of passive galaxies in clusters is responsible for 
this effect since, as we show, old passive galaxies host SNe Ia with smaller stretch 
than young passive galaxies ($\sim 4\sigma$).
\end{abstract}


\section{Introduction}
\label{sec:Introduction}

Type Ia supernovae (SNe Ia) have been an important cosmological tool as distance
indicator, being used to constrain the acceleration of the universe 
\citep{Perlmutter98mn,Riess98mn}, especially after the establishment of 
relations between their light-curve shape, their colour and their 
absolute magnitude at peak \citep{Phillips93mn,Riess96mn}. These relations
allow us to measure the luminosity distance with an average $\sim0.15$
magnitude precision up to redshifts $z\sim 1$ \citep{Conley11mn}. 

In order to improve these distance measurements, considerable attention
has been dedicated to develop and validate the standardisation
of type Ia supernova luminosities, and recent studies have supported
its correlation with host galaxy properties, spectral features and
flux ratios \citep{Kelly10mn,Bailey09mn,Chotard11mn}. Regarding the 
environmental influence on SNe Ia characteristics, many authors have 
recently reported that different galaxies host slightly different SNIa populations,
and that accounting for this preference can further increase distance
measurements precision \citep{Hamuy95mn,Riess99mn,Hamuy00mn,Sullivan06mn,
Gallagher08mn,Sullivan10mn,Lampeitl10mn,Gupta11mn,DAndrea11mn}.
This is likely to be an important issue for precise distance measurements 
in cosmology since galaxy population changes with redshift.

An example of such reports is given by \citet{Lampeitl10mn}, who analysed 
low redshift ($z<0.21$) data from the Sloan Digital Sky Survey-II 
\citep[SDSS-II,][]{York00mn, Frieman08mn} separating the SNe Ia by 
their host galaxy specific star formation rate, which was derived 
from photometry. Using this method, they showed that passive galaxies
tend to host SNe Ia that are in many ways different from 
their counterparts in star-forming galaxies: 
(1) passive galaxy SNe Ia have faster-declining light curves; 
(2) the correlation between their colour and their luminosity is weaker; 
(3) after correcting for their colour and light-curve shape 
(where the colour correction is different from the other SNe Ia), 
they are intrinsically brighter by $\sim0.1$ mag and their Hubble Residuals 
present less scatter.
\citet{Sullivan10mn} analysed the Supernova Legacy Survey 
\citep[SNLS,][]{Astier06mn} data up to higher redshifts using 
host galaxy mass derived from photometry, and demonstrated that SNe Ia in massive
hosts tend to have similar properties to the ones described above. 
\citet{Hicken09mn} used the host galaxy morphology
and found evidence that E/S0 galaxies tend to host brighter SNe Ia than
Scd/Sd/Irr galaxies. \citet{DAndrea11mn} analysed
spectra from low redshift ($z<0.15$) host galaxies and found that
SNe Ia in high-metallicity hosts are $\sim0.1$ magnitudes brighter
than those in low-metallicity hosts (after light curve correction).
The variety of methods and databases used in all those works indicate 
that the results are robust. 
In contrast, differences in SNe Ia colour 
have been more elusive. While \citet{Lampeitl10mn} could not identify any 
differences between colours of SNe Ia in active and passive hosts, 
\citet{Sullivan10mn} found weak evidences that passive galaxies host 
bluer SNe Ia than active galaxies, whereas \citet{Gupta11mn} found 
that older galaxies may host redder SNe Ia, apparently an opposite result. 
\citet{Smith12mn} addresses this issue and shows that 
this result might depend on a more precise classification of the hosts.

The dependence of SNIa properties on environment is also important for 
the study of SNIa rates, both for SNe in different host galaxy types 
and for SNe inside and outside galaxy clusters, since different properties
can lead to different selection effects. Supernovae are a major source
of metal enrichment for galaxies and clusters, and their rates and
properties are crucial to constrain possible enrichment processes 
\citep[e.g.][]{Domainko04mn}. The study of SNIa rates and their delay 
time distribution (DTD) have also indicated the existence of two different 
populations, called `delayed' and `prompt' types, and papers on SNIa 
properties have correlated DTD with other properties such as light-curve 
stretch \citep{Mannucci06bmn, Sullivan06mn, Smith12mn}. Further understanding 
of this relation will require good assessment of variations observed in 
SNIa properties.

SNe Ia in clusters are also particularly interesting. Since the work of 
\citet{Zwicky51mn}, it has been suspected that galaxy clusters possessed 
a population of intergalactic, free-floating stars which were probably 
torn from their host galaxies by tidal forces. Such stars could 
lead to hostless intracluster supernovae, and direct detection of these 
stars and supernovae were reported by \citet{Ferguson98mn} and 
\citet{GalYam03mn}, respectively. These SNe could present different 
properties from their intragalactic counterparts (for instance, due to 
the absence of host dust extinction). In addition, cosmological SNIa 
surveys which target clusters specifically \citep[e.g.][]{Dawson09mn} 
may require a thorough understanding of such objects to avoid potential 
biases. 

Primarily because of the lack of large enough samples, there
have been no published investigations on property differences between SNe Ia
inside and outside galaxy clusters. Papers that analysed SNe Ia inside
clusters have been able to amass from 1 to 27 objects and focused
on determining their rate \citep{GalYam02mn,Graham08mn,Mannucci08mn,Dilday10mn}.

By making use of a larger galaxy cluster catalogue, the Gaussian Mixture 
Brightest Cluster Galaxy \citep[GMBCG,][]{Hao10mn}, 
and a larger photometrically-typed supernova 
sample possessing host galaxy spectroscopic redshifts (\emph{spec-z}) 
from the Baryon Oscillation Spectroscopic Survey 
\citep[BOSS,][]{Eisenstein11mn,Dawson13mn}, we present 
the first study on the properties of SNe Ia residing in rich 
galaxy clusters. Here we searched for possible
statistical differences in SNIa parameters and in their correlation
with host galaxy properties (derived from photometry spanning the
ultraviolet, optical and near infrared bands) when comparing SNe Ia
inside and outside GMBCG clusters. This work contributes to the 
study of SNIa rates, SNIa physics and of systematic effects on 
distance measurements.

This paper is organised as follows: in section \ref{sec:dataset} we 
describe the SNIa data, the galaxy cluster catalogue, the BOSS 
spectroscopic survey and the galaxy photometry used in this work; 
in \ref{sub:sn-model-fitting} and \ref{sec:host-galaxies} we present 
the methods used for fitting models and extracting parameters to SNe Ia 
and galaxies, along with the host galaxy identification method; 
in \ref{sub:cluster-sn-selection} we introduce our method for identifying 
SNe Ia residing in clusters and present a few crosschecks; and in 
\ref{sub:comparing-samples} we describe our method for comparing different SNIa samples. 
In section \ref{sec:known-relations} we recover known relations between SNe Ia and 
their hosts in order to validate our analysis and compare our results regarding the 
cluster SNe Ia, which are presented in section \ref{sec:cluster-sn-properties}. Section \ref{sec:host-age} 
compares SNe Ia hosted by young and old passive galaxies. We verify in \ref{sec:rubustness-tests} 
how our results are affected by differences in our procedures, and in \ref{sec:cluster-radius}, 
in particular, the influence of a smaller angular separation between the supernova and the cluster centre. 
We conclude and summarise our findings in section \ref{sec:conclusions}. 
Appendix \ref{sec:Contamination-estimation} gives details about our cluster SNIa selection, 
and Table \ref{tab:fulltable} presents the complete dataset for our cluster SNIa sample.

\section{Dataset}
\label{sec:dataset}

\subsection{Supernovae}

The supernovae dataset used in this work was obtained by the Sloan Digital 
Sky Survey-II (SDSS-II) Supernova Survey over the region 
of the sky called Stripe 82, an equatorial
stripe with declination $-1.26^{\circ}<\delta<+1.26^{\circ}$ and right
ascension $-60^{\circ}<\alpha<+60^{\circ}$ \citep{York00mn, Frieman08mn}. 
The Stripe 82 was imaged on all \emph{ugriz} filters every 4 days, on 
average, during the fall seasons of 2005--2007. The image processing pipeline and 
transient selection criteria is presented in \citet{Sako08mn}. 
The camera and photometric system used for collecting the data are described 
in \citet{Gunn98mn} and \citet{Fukugita96mn}. 
This SNe dataset contains 504 spectroscopically confirmed SNe Ia and 752 SNe
photometrically typed as Ias with \emph{spec-z} of their hosts 
\citep[in prep.]{Sako08mn,Holtzman08mn, Sako12mn},
making a total of 1256 SNe Ia with spectroscopic redshifts.

The supernovae
were photometrically typed using the {\sc psnid} software \citep{Sako11mn}. 
We did not make use of supernovae with only photometric redshifts
because of the high contamination by type Ibc SNe, and we expect
a $\sim 5$ per cent contamination by different SN types (especially Ibc) 
in the photometrically typed SNe Ia with \emph{spec-z} \citep[in prep.]{Sako12mn}, 
resulting in $\sim 4$ per cent contamination for the whole sample. 

Before fitting the light curves for its parameters, the following 
quality cuts were required from the data:

\begin{enumerate}
\item a minimum of 5 different observed epochs;
\item at least one observation after the light-curve peak;
\item at least one observation before 5 days after the light-curve peak, 
in the SN rest-frame;
\item at least two observations in different filters with signal-to-noise 
ratio (SNR) greater than 4. 
\end{enumerate}
These cuts reduced the number of SNe Ia to 451 spectroscopically and 679 
photometrically typed. We also removed from the remaining ones 8 
spectroscopically typed SNe Ia known to be peculiars, making a total of 
1122 SNe Ia. Tighter constraints on light-curve measurements like those 
employed for cosmology fitting \citep[e.g.][]{Kessler09bmn} were not used 
in order to maximize the amount of SNe Ia in our samples, although some 
extra cuts were applied after light curve fitting to minimize contamination 
and to remove outliers (sections \ref{sub:sn-model-fitting} and \ref{sub:x1-c-cuts}).  

The vast majority (87 per cent) of the host \emph{spec-z} used in this work was measured by the 
BOSS project and its SNe host galaxy ancillary program \citep{Dawson13mn, Bolton12mn}, 
while the remaining was measured by the SDSS-II SN survey spectroscopic follow-up program \citep{Frieman08mn}.  
BOSS is a part of SDSS-III collaboration \citep{Eisenstein11mn} aimed at measuring 
the redshift of 1.5 million luminous galaxies up to $z\sim 0.7$ and over 100,000 
$z\sim 2$ quasars using a 1,000 fiber spectrograph mounted on the Sloan Foundation 
2.5-meter telescope at Apache Point Observatory \citep{Gunn06mn,Smee12mn}. The 
project is expected be completed on 2014 and the latest data release (DR9) 
presented the spectra of 535,995 galaxies \citep{Ahn12mn}. Its host galaxy ancillary 
program is already finished. For a description of the BOSS target selection for 
supernovae hosts, see \citet[in prep.]{Campbell13mn,Olmstead13mn}.   

\subsection{Galaxy clusters}

The galaxy clusters employed here were identified using the GMBCG 
algorithm \citep{Hao10mn} on the SDSS DR7 \citep{Abazajian09mn} galaxy catalogue. 
Only photometric information was used. The GMBCG algorithm relies on a typical 
charactistic of galaxy clusters: the presence of a number of galaxies with 
similar colour -- called `red sequence galaxies' -- accompanied by a very bright, 
central galaxy, called `brightest cluster galaxy' (BCG).

At a glance, the algorithm starts by selecting a bright galaxy from the 
galaxy catalogue (a potential BCG). Then, it selects 
all fainter galaxies within a projected 0.5 Mpc radius from the candidate BCG that 
fall inside a broad photo-z window ($\pm0.25$) around it. Finally, it fits 
these galaxies colour distribution with two 
Gaussians. If the candidate BCG is likely to belong to the reddest Gaussian and the 
latter is sufficiently narrow (thus a potential red sequence), then a galaxy cluster 
centre is identified at the BCG's position, unless this galaxy can be selected as member 
of a different, denser cluster. 

To estimate the clusters richness $r$ (i.e., number of member galaxies), the algorithm 
basically counts the number of galaxies $N_{\mathrm{gals}}$ that: 
(1) are brighter than $0.4L^*$ and dimmer than the BCG ($L^*$ is the characteristic luminosity in the Schechter luminosity function); 
(2) can be considered members of the red sequence;
and (3) sit inside a radius around the BCG in which the estimated density 
is $\sim 200$ times the critical density. 
Only clusters with $N_{\mathrm{gals}}\geq 8$ were included in the 
public catalogue we used; these systems are termed `rich clusters'. 
The purity and completeness of the catalogue were estimated for various 
bins of richness and redshift by applying the GMBCG algorithm to a mock 
catalogue. Its completeness is greater than 90 per cent for all bins and its purity 
ranges from $\sim 60$ per cent for clusters with $N_{\mathrm{gals}}=10$ to 
$\sim 90$ per cent or more for clusters with $N_{\mathrm{gals}}\geq 15$. 
For more details on the cluster catalogue construction and 
characteristics, see \citet{Hao10mn}.

\begin{figure}
\includegraphics[width=1\columnwidth]{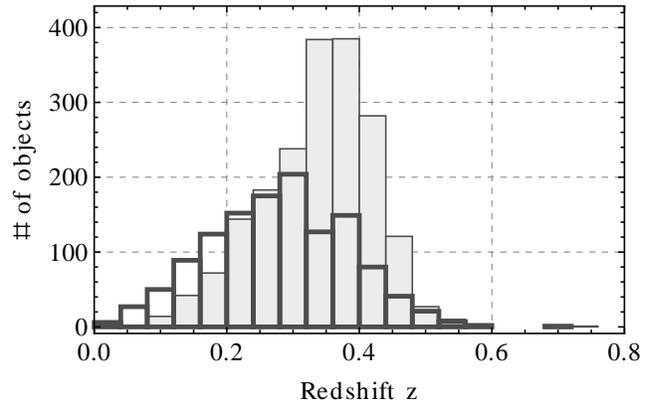}
\caption{Histogram for the 1256 supernovae type Ia without any cuts 
(thick contours, no filling) and the 1905 GMBCG clusters (thin contours, gray filling) 
used in this paper.}
\label{fig:Objects_z_Histogram}
\end{figure}

The SDSS GMBCG cluster catalogue includes the Stripe 82 region, where 1905 
rich clusters were identified. All clusters in the catalogue 
have a photometric redshift, and 576 of the clusters in the Stripe 82 region have
the spectroscopic redshift of its BCG. A redshift distribution histogram
for both the SNe Ia and the GMBCG clusters is presented in Fig. 
\ref{fig:Objects_z_Histogram}.

\subsection{Host galaxy photometry}

The search for the SNe Ia's host galaxy was done exclusively 
in the SDSS DR8 \citep{Aihara11mn} primary objects list, which 
includes the highest quality SDSS runs over the Stripe 82 region.  
Within the Stripe 82, approximately 5 million galaxies were detected. 
The evaluation of host galaxy properties was entirely photometric. 
When available, we supplemented SDSS DR8 photometry with ultraviolet
and near infrared measurements taken by the Galaxy Evolution Explorer
\citep[\emph{GALEX},][]{Martin05mn} General Release 6 (GR6) and the UKIRT Infrared Deep Sky Survey
\citep[UKIDSS,][]{Lawrence07mn} Data Release 8 (DR8). \emph{GALEX} has filters in far-UV and near-UV
bands, while UKIDSS has filters in the YJHK bands \citep{Hewett06mn}. 
We present our methods for identifying a SNIa's host galaxy and 
for estimating its properties in section \ref{sec:host-galaxies}.

\section{Methodology}

\subsection{SNIa model fitting}
\label{sub:sn-model-fitting}

All supernovae type Ia were fitted using the SALT2 model 
\citep{Guy07mn} implemented by the publically available software
{\sc snana} \citep{Kessler09amn}. This SNIa light-curve model is
based on five parameters: the redshift $z$, the time of
maximum $T_{0}$, a overall normalisation $x_{0}$, a stretch
or `light-curve width' parameter $x_{1}$
and a colour parameter $c$. The normalisation $x_{0}$ is related
to the apparent magnitude at peak in the \emph{B}-band $m_{B}$ by:

\begin{equation}
m_{B}=10.635-2.5\log_{10}x_{0}\;.
\label{eq:mB_relation}
\end{equation}
The distance modulus $\mu\equiv5\log_{10}(\frac{d_{\mathrm{L}}}{10\;\mathrm{pc}})$, 
where $d_{\mathrm{L}}$ is the luminosity distance, 
is calculated using corrections based on $x_{1}$ and $c$ that account
for the fact that SNe Ia with wider light curves ($x_{1}>0$) tend to
be brighter, and redder SNe Ia ($c>0$) tend to be dimmer:

\begin{equation}
\mu=m_{B}-M+\alpha x_{1}-\beta c\;.
\label{eq:distmod_relation}
\end{equation}
In the equation above, $M$ (an average absolute magnitude), $\alpha$
and $\beta$ (often called `nuisance parameters') are obtained 
from a sample of SNe Ia so that the $\chi^{2}$ for $\mu$ around the best-fit
cosmology is minimized. When calculating $\chi^{2}$, in addition to the 
measurement errors, we included an intrinsic dispersion $\sigma_{\mathrm{int}}$ 
such that the minimum reduced $\chi^{2}$ is set to $\sim 1$, as commonly done by 
papers that use the SALT2 model \citep[e.g.][]{Sullivan10mn, Lampeitl10mn, 
Campbell13mn}.

For the determination of $\mu$, $M$, $\alpha$ and $\beta$ the software {\sc salt2mu} 
in {\sc snana} package was used \citep{Marriner11mn}. While searching for the best 
$M$, $\alpha$ and $\beta$ values, instead of varying the cosmological density parameters 
for matter and dark energy ($\Omega_m$ and $\Omega_\Lambda$) or the dark energy equation 
of state, {\sc salt2mu} uses a constant fiducial cosmological model and parametrizes 
deviations from it due to cosmology and other redshift dependent effects with different 
SNe Ia absolute magnitudes at different redshift ranges. Thus, even though we adopted 
throughout this work a fiducial flat $\mathrm{\Lambda CDM}$ cosmological model with 
$\Omega_{\mathrm{m}}=0.27$, $\Omega_{\mathrm{\Lambda}}=0.73$ and $H_{0}=70\;\mathrm{km\;s^{-1}\;Mpc^{-1}}$ 
\citep{Kessler09bmn}, our nuisance parameters are not constrained by this particular 
model.

The Hubble residuals ($HR\equiv\mu_{\mathrm{SN}}-\mu_z$) are given by the difference 
between the distance modulus $\mu_{\mathrm{SN}}$ obtained from data via Eq. 
\ref{eq:distmod_relation} and the expected distance modulus $\mu_z$ from our fiducial 
cosmology, thus they are the residuals \emph{after} correction by colour and stretch. 

After fitting the SNe Ia light curves, we removed 12 outliers from our 1122 SNe Ia based 
on their light curve properties: 9 with $|x_1|\geq 5$ and 3 extremely red SNe Ia with $c>0.45$. 
Furthermore, we excluded 37 SNe Ia whose SALT2 fit $\chi^{2}$ probability was smaller than 0.01. 
Lastly, we removed 10 outliers that were unusually off the Hubble diagram (more that 4$\sigma$), 
reducing our sample to 414 spectroscopically and 649 photometrically typed SNe Ia. 
Even for a large sample, these outliers alone can alter significantly the nuisance parameters and 
the intrinsic scatter, and we are interested in values that are representative of the whole sample.

\subsection{Host galaxies}
\label{sec:host-galaxies}

\subsubsection{Identification of the host galaxy}
\label{sub:host-identification}

The identification of a SNIa's host galaxy was done by searching in the 
SDSS DR8 primary objects list for all galaxies within a 30 arcsec 
radius of the SNIa. We then selected as host the galaxy whose angular 
separation from the SNIa, normalised by the angular elliptical radius of 
the galaxy in the direction of the SNIa (called `directional light 
radius'), $d_{\mathrm{DLR}}$, was the smallest. 
To compute the elliptical radius we used the Petrosian half-light 
radius as a measure of the size of the galaxy and its Stokes parameters $Q$ 
and $U$ as a measure of its ellipticity and orientation, all in the $r$ band 
\citep{Abazajian09mn}. When these parameters were unavailable the object 
in question was not considered a viable host. To avoid 
misidentifications we also imposed a maximum $d_{\mathrm{DLR}}$ of 4. 

Of the 1063 SNe Ia selected in section \ref{sub:sn-model-fitting}, 1017 
(96 per cent) have an associated host. More information about the host 
identification process is available on the SDSS-II Supernova Survey 3-Year 
Data Release paper \citep[in prep.]{Sako12mn}. A similar process of host 
identification was adopted in \citet{Sullivan06mn}.

\subsubsection{Host galaxy properties}
\label{sub:host-properties}

The host galaxy properties were estimated by fitting synthetic spectral 
energy distributions (SED) to the galaxy photometry obtained by the SDSS, 
\emph{GALEX} and UKIDSS surveys. 
The matching among the surveys was done by selecting the object in
UKIDSS and/or \emph{GALEX} catalogues nearest (on the sky plane) to a SDSS galaxy, with
a maximum angular separation of $5$ arcsec. Out of 1017 SNe Ia with identified SDSS host, 
455 had matches in both \emph{GALEX} and UKIDSS, 222 had a match only in UKIDSS, 239 only 
in \emph{GALEX} and 101 had no matches in both catalogues. 
The magnitude measurements used were Model magnitudes for SDSS 
\citep{Stoughton02mn}, Petrosian for UKIDSS and Kron-like elliptical
aperture magnitude for \emph{GALEX} \citep{Petrosian76mn,Kron80mn}. 
More information about our methods for combining photometry can be 
found in \citet{Gupta11mn}.

To generate the synthetic SEDs we used the Flexible Stellar Population 
Synthesis (FSPSv2.1) code \citep{Conroy09mn,Conroy10mn},
with the same procedure as in \citet{Gupta11mn} (with the sole difference
in the cosmological parameters used). The basic inputs were the
stellar spectral library BaSeL3.1 \citep{Lejeune97mn,Lejeune98mn}, the Padova stellar
evolution model \citep{Marigo07mn,Marigo08mn}, the
Initial Mass Function (IMF) from \citet{Chabrier03mn} and the dust
model from \citet{Charlot00mn}. For more details,
please refer to \citet{Conroy09mn}.

The SEDs were generated on a
grid of four FSPS parameters: the time when star formation begins 
$t_{\mathrm{start}}$; a star formation rate (SFR) time scale 
$\tau_{\mathrm{SF}}$, where $\mathrm{SFR}(t)\propto e^{-t/\tau_{\mathrm{SF}}}$; 
the metallicity $\log(Z/Z_{\odot})$, assumed constant over time; and
a coefficient $\tau_{\mathrm{dust}}$ for the optical depth $\tau(t)$ around
the stars of age $t$, given by:
\begin{equation}
\tau(t)=\begin{cases}
3\tau_{\mathrm{dust}}\left(\frac{\lambda}{5500\mathrm{\AA}}\right)^{-0.7}\,, & t\leq10\;\mathrm{Myr}\,,\\
\tau_{\mathrm{dust}}\left(\frac{\lambda}{5500\mathrm{\AA}}\right)^{-0.7}\,, & t>10\;\mathrm{Myr\,.}
\end{cases}
\label{eq:dust_optical_depth}
\end{equation}
The model fluxes on the far-UV, near-UV, \emph{ugriz} and \emph{YJHK} bands
were then calculated using these SEDs. 
The measured fluxes were corrected for galactic extinction using the
Cardelli curve \citep{Cardelli89mn} and Milky Way dust maps \citep{Schlegel98mn}, and 
the SDSS and UKIDSS magnitudes were corrected to the AB system using \citet{Kessler09bmn} 
and \citet{Hewett06mn}, respectively. The best fit model was chosen by comparing these 
measured fluxes to the model fluxes using the least squares method. No requirements were 
made regarding the number of bands measured.

Three host galaxy properties were estimated from the fits: the stellar
mass (amount of mass in the form of stars), the mass-weighted average
age and the specific star formation rate (sSFR). The stellar mass
was obtained by multiplying the de-reddened measured \emph{r} band
luminosity with the model mass-to-light ratio on the same band. The
mass-weighted average age was calculated as:

\begin{equation}
\langle\mathrm{Age}\rangle=A-\frac{\int_{0}^{A}t\Psi(t)dt}{\int_{0}^{A}\Psi(t)dt}\,,
\label{eq:def_host_age}
\end{equation}
where $A$ is the age of the universe at the galaxy's redshift minus
$t_{\mathrm{start}}$ and $\Psi$ is the SFR. The sSFR was obtained by normalising
$\Psi(t)$ over the period $A$ to unity, and taking the average 
over the interval $A-250\;\mathrm{Myr}<t<A$. To reduce the amount of 
noise in the host property analysis, we disconsidered hosts 
whose photometry fit presented a chi-square $p$-value smaller than 
0.001. This reduced the number of available hosts from 1017 to 
717, mainly due to the models in our grid being non-representative.\footnote{
The number of available hosts only limits our SNe Ia sample sizes when 
host information is required. Otherwise, the full SNe Ia sample (1063) is used.} 
Part of these exclusions may also be caused by matching the wrong objects through 
\emph{GALEX}, UKIDSS and SDSS catalogues. A detailed description of the 
host galaxy's properties estimation can be found in \citet{Gupta11mn}.

When necessary, we separated our hosts in two groups based on their 
sSFR ($\mathrm{sSFR}<-11.72$ are called `passive' and $\mathrm{sSFR}>-10.5$, 
`active'). These sSFR limits were based on \citet{Lampeitl10mn} -- the 
-11.72 limit for passive galaxies is, on average, 1$\sigma$ below the 
active galaxies limit of -10.5 -- and were chosen so the separation between these 
two classes is clean. 438 hosts galaxies were classified as active and 162 were 
classified as passive.

\subsection{Selection of SNe Ia as members of clusters}
\label{sub:cluster-sn-selection}

With the purpose of classifying a SNIa as a member of a galaxy cluster we defined 
three criteria that should be fulfilled: their angular positions should be compatible, 
their redshifts should be compatible, and the cluster in the catalogue should be real and not 
a projection of field galaxies. For the last two criteria we adopted a probabilistic approach 
which combined them into a single condition described by Eq. \ref{eq:prob-product}. 
These criteria are described below in detail.

\subsubsection{Selection of SNe Ia projected on clusters}
\label{sub:projection-test}

In order to identify the SNe Ia that are inside SDSS GMBCG galaxy clusters,
we started by selecting all supernovae type Ia within a projected
$1.5\mathrm{Mpc}$ physical radius around any cluster, as done in
previous studies of SNIa rate in galaxy clusters \citep[see][]{Mannucci08mn,Dilday10mn}.
For a given SNIa $s$ and a cluster $k$, this selection translates
into obeying the following relation:

\begin{equation}
\cos\delta_{s}\cos\delta_{k}\cos\left(\alpha_{s}-\alpha_{k}\right)+\sin\delta_{s}\sin\delta_{k}
\geq\cos\left(\theta_{\mathrm{max}}^{(k)}\right)\;,
\label{eq:cosep}
\end{equation}
\begin{equation}
\theta_{\mathrm{max}}^{(k)}\equiv\frac{1.5\mathrm{Mpc}(1+z_{k})}{c\int_{0}^{z_{k}}\frac{dz}{H(z)}}\;,
\end{equation}
where $\theta_{\mathrm{max}}^{(k)}$ is the angular radius of the cluster $k$,
$c$ is the speed of light, $\alpha_{s}$ and $\delta_{s}$ are the right 
ascension and declination of the SNIa, $\alpha_{k}$, $\delta_{k}$ and $z_{k}$ are the right ascension,
declination and redshift of the cluster, respectively, and $H(z)$ is the 
Hubble parameter, given by: 

\begin{equation}
H(z) = H_0 \sqrt{\Omega_{\mathrm{m}}(1+z)^3+\Omega_{\mathrm{\Lambda}}}\;.
\label{eq:Hubble-Parameter}
\end{equation}

Of the 414 spectroscopically confirmed SNe Ia, 82 are projected onto clusters 
(21 of these on more than one); and of the 649 photometrically typed SNe Ia, 
148 are projected, 32 being onto more than one cluster.

\subsubsection{Redshift compatibility}

The next step for determining if a SN belongs to a cluster was 
to check for redshift compatibility between the SN and the clusters
onto which they were projected. Since galaxy clusters are gravitationally
bound objects, there is no Hubble flow inside them, and if it were
not for peculiar velocities of its members, all objects inside it
would have the same redshift. Therefore, the tolerance on redshift
difference between the supernova and the cluster arises from a combination of the velocity
dispersion inside the cluster, which we assumed to be $\sigma_v=500$ $\mathrm{km\;s^{-1}}$, 
and measurement errors.

For each pair of cluster and projected SNIa we calculated the probability $p$ 
for their redshift difference to be inside a characteristic
range. We assumed that the supernova and cluster redshift probability distribution
were Gaussians $N(z_{s},\sigma_{s})$ and $N(z_{k},\sigma_{k})$,
respectively, where $z_{s}$ and $\sigma_{s}$ ($z_{k}$ and $\sigma_{k}$)
are the redshift assigned to the supernova (cluster) and its uncertainty.
The probability distribution for the difference in redshift is then
$N(z_{s}-z_{k},\;\sqrt{\sigma_{s}^{2}+\sigma_{k}^{2}})$, and the
probability for compatible redshifts was calculated as:

\begin{equation}
p=\frac{1}{\sqrt{2\pi (\sigma_{s}^{2}+\sigma_{k}^{2})}}\int_{-z_{\mathrm{d}}}^{z_{\mathrm{d}}}e^{-\frac{[z-(z_{s}-z_{k})]^2}{2(\sigma_{s}^{2}+\sigma_{k}^{2})}}dz\;.
\label{eq:compatzspec}
\end{equation}

The choice of $z_{\mathrm{d}}$ depended on the type of redshift assigned to the cluster. 
For the 576 clusters with spectroscopically confirmed BCGs, the BCG \emph{spec-z} was used, 
in which case the maximum redshift difference was $z_{\mathrm{d}}=0.005$, 
corresponding to a maximum velocity difference of $3\sigma_v=1500$ $\mathrm{km\;s^{-1}}$. 
For the 1329 remaining clusters with photometric redshifts only, the choice 
was $z_{\mathrm{d}}=0.030$. The process for choosing these values is described in Appendix 
\ref{sec:Contamination-estimation}. No distinction was made on whether 
$z_{s}$ was the supernova's redshift itself or its host galaxy's redshift.

\subsubsection{Cluster existence and final selection}

For all clusters with SNe Ia projected onto them 
we calculated the probability $q$ of it being truly a cluster and not just a projection 
of field galaxies. We assumed that such probability is equal to the 
purity estimated by \citet{Hao10mn} for the cluster catalogue in the 
redshift and richness range accessed by the cluster in question. These 
are presented in Table \ref{tab:cluster-purity}.
  
\begin{table}
\caption{Probability $q$ that a cluster in the redshift range given by 
the first column and in the richness range given by the first row is 
real. The values are based on the purity estimations made using mock catalogues.}
\label{tab:cluster-purity}
\begin{tabular}{|r|c|c|c|c|}
\hline
              & $r<15$ & $15\leq r<20$ & $20\leq r<25$ & $r\geq 25$ \tabularnewline
$z<0.15$      & 0.78   & 0.96          & 1.00           & 0.99       \tabularnewline 
$0.15<z<0.20$ & 0.70   & 0.92          & 0.98           & 0.98       \tabularnewline
$0.20<z<0.25$ & 0.70   & 0.89          & 0.98           & 0.98       \tabularnewline
$0.25<z<0.30$ & 0.55   & 0.82          & 0.92           & 0.96       \tabularnewline
$0.30<z<0.35$ & 0.48   & 0.81          & 0.91           & 0.95       \tabularnewline
$0.35<z<0.40$ & 0.62   & 0.89          & 0.92           & 0.98       \tabularnewline
$0.40<z<0.45$ & 0.51   & 0.84          & 0.93           & 0.94       \tabularnewline
$z>0.45$      & 0.78   & 0.90          & 0.90           & 0.97       \tabularnewline
\hline
\end{tabular}
\end{table}

The final step for selecting SNe Ia as cluster members was to pick from 
the projected ones those obeying the relation:

\begin{equation}
qp\geq P_{\mathrm{min}}\;,
\label{eq:prob-product}
\end{equation}
where $P_{\mathrm{min}}=0.5$ is a minimum probability chosen through the 
procedure described in Appendix \ref{sec:Contamination-estimation}. 
This equation states that a SNIa is only considered to be inside a cluster 
if the cluster is real and their redshifts are compatible. Assuming that 
these conditions are independent, the probability of fulfilling both 
criteria is equal to the product of $q$ and $p$.

After selecting for redshift compatibility with a real cluster, 6 (15) spectroscopically
confirmed and 17 (10) photometrically typed SNe Ia were assigned to 
clusters with spectroscopic (photometric) redshifts, 
making a total of 48 cluster type Ia supernovae, which are 
presented in Table \ref{tab:fulltable}. As explained in Appendix 
\ref{sec:Contamination-estimation}, the contamination by field SNe Ia was 
estimated as 29 per cent. The 1015 SNe Ia that did not pass this selection 
criteria were considered to be field SNe Ia. Since the small size of the 
cluster SNe Ia sample and its high contamination will dominate the noise 
during sample comparisons, more strict cuts on the field sample are 
unnecessary for our purposes. 
Table \ref{tab:summary-numbers} summarises the number of SNe Ia obtained 
after each cut and cluster sample selection step were taken.

\begin{table}
\caption{Amount of SNe Ia selected in each step of the cluster sample formation and 
its sub-divisions by host galaxy type. The columns present, from left to right, 
the number of SNe Ia: spectroscopically typed; photometrically typed; total (sum of 
the two);  and left outside the cluster sample (therefore, in the field sample). 
Apart from the last one, all lines are a sub-set of the previous one, and show the 
number of SNe that: were typed as Ias by {\sc psnid}; passed light-curve cuts; were 
not known peculiars; had $|x_1|<5$ and $|c|<0.45$; their SALT2 fit $\chi^2$ probability 
was larger than 0.01; are not outliers in the Hubble residuals; are projected onto clusters; 
were selected as cluster members; has an identified host; its host fit passed the chi-square 
test; its host was classified as active; its host was classified as passive.}
\label{tab:summary-numbers}
\begin{tabular}{|l|c|c|c|c|c|}
\hline
                       & Spec. & Phot. & Total & Field total \tabularnewline\hline
Initial                & 504   & 752   & 1256  & --          \tabularnewline 
LC cuts                & 451   & 679   & 1130  & --          \tabularnewline
No peculiars           & 443   & 679   & 1122  & --          \tabularnewline
$x_1$--$c$ cuts        & 439   & 671   & 1110  & --          \tabularnewline   
$P(\chi^2)>0.01$       & 414   & 659   & 1073  & --          \tabularnewline
$HR$ $4\sigma$ cut     & 414   & 649   & 1063  & --          \tabularnewline
Projected              & 82    & 148   & 230   & --          \tabularnewline
$qp\geq P_{\mathrm{min}}$ & 21    &  27   & 48    & 1015        \tabularnewline
w/ host                & 19    &  27   & 46    & 971         \tabularnewline
w/ host fit            & 11    &  21   & 32    & 685         \tabularnewline
Active                 & 2     &  7    &  9    & 429         \tabularnewline
Passive                & 7     &  12   & 19    & 143         \tabularnewline
\hline
\end{tabular}
\end{table}

\subsubsection{Selection crosschecks}
\label{sub:Selection-crosschecks}

We compared our supernova selection with the one done by \citet{Dilday10mn} 
(hereafter, D10) using the maxBCG cluster catalogue \citep{Koester07mn}. 
In this work we made use of BOSS redshifts
for supernova hosts, an option not available for D10. 
This advantage drastically increased the number of SNe Ia with spectroscopic
redshifts (by 652) and its precision, resulting in better typing
and, in particular, better redshift comparison with galaxy clusters. 
Whereas the maxBCG cluster catalogue used the SDSS DR4 \citep{Adelman-McCarthy06mn} 
and only one colour to find the red sequence, GMBCG used 
DR7 and two colours, thereby increasing the redshift depth and the number
of clusters detected in the Stripe 82 region from 492 to 1905. This 
changes are expected to make our cluster SNIa sample larger, and thus 
should include the majority of supernovae selected by D10. However, 
D10 only accounted for cluster contamination by field galaxy projections 
when calculating SNIa rates and not during SNIa selection. Furthermore, they did not 
eliminate outliers based on light-curve parameters. Therefore, a few 
D10 SNe Ia will be excluded by our method. The comparison can also be 
affected by differences in goal and methodology. 

D10 found 27 SNe Ia in maxBCG clusters, 6 of which
were not used in this work because of lack of \emph{spec-z} or differences
in the SN typing. Out of the remaining 21, 11 were also selected by
our criteria, 2 would have been selected if we had not accounted for cluster catalogue 
contamination and 1 was eliminated because of its colour of $0.73\pm0.07$. 
From the remaining 7, 2 did not have a projected GMBCG cluster,
3 were excluded because of differences in the cluster redshift, and
2 because of more restrictive requirements used for the redshift compatibility.

We found 37 cluster SNe Ia not selected by Dilday, 14 of which are beyond
maxBCG redshift limit of 0.3. From the remaining 23, 6 had their redshifts 
measured only by BOSS, 8 were not projected onto a maxBCG cluster and 1 was 
projected on a cluster with a different redshift. The last 8 SNe Ia had 
compatible maxBCG clusters and the reasons why they were not included in D10 
could not be traced. It is possible that they were eliminated as non-Ias or 
were excluded by data quality cuts. Since the majority of the differences between 
D10 sample and ours are due to SN typing and the cluster catalogue used, we considered 
our selection methods compatible.

\begin{figure}
\includegraphics[width=1\columnwidth]{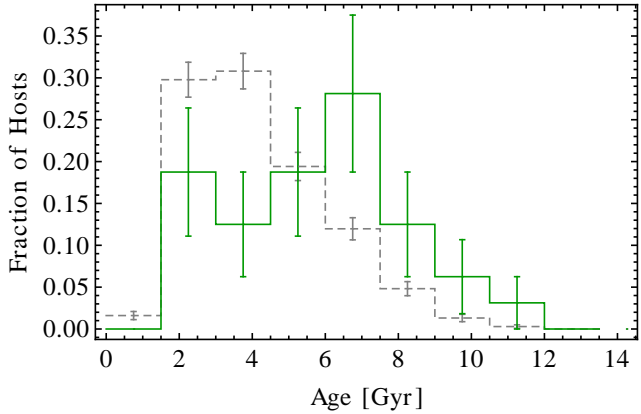}
\caption{Histogram of the host galaxies mass-weighted average age for 32 cluster SNe Ia
(green, solid line) and 685 field SNe Ia (gray, dashed line) that had good host fits. 
The cluster sample hosts are, on average, older than the hosts in the field sample.}
\label{fig:age_hist}
\end{figure}

\begin{figure}
\includegraphics[width=1\columnwidth]{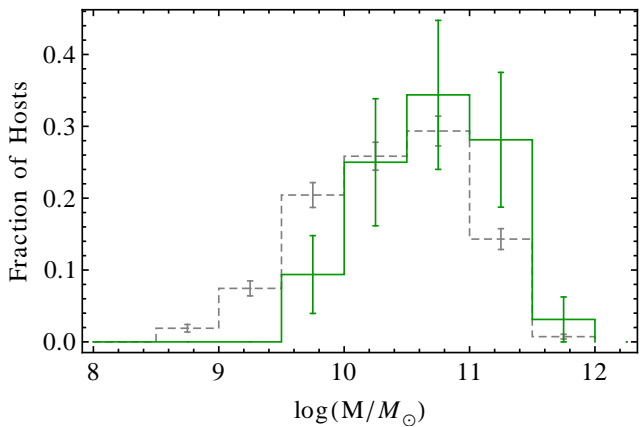}
\caption{Histogram of the host galaxies stellar mass for 32 cluster SNe Ia
(green, solid line) and 685 field SNe Ia (gray, dashed line) that had 
good host fits. The cluster sample host mass distribution is shifted to 
larger values.}
\label{fig:mass_hist}
\end{figure}

\begin{figure}
\includegraphics[width=1\columnwidth]{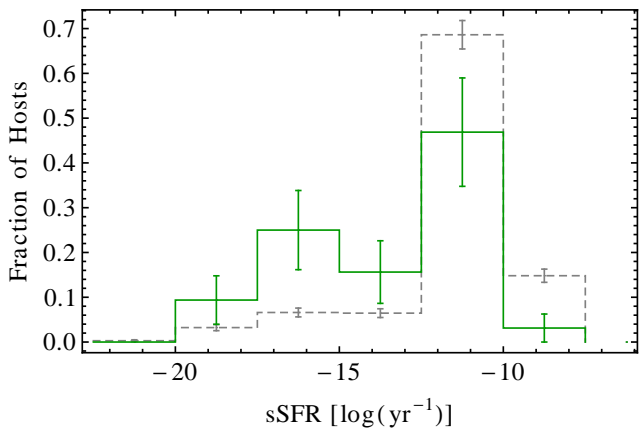}
\caption{Histogram of the host galaxies specific star formation rate 
for 32 cluster SNe Ia (green, solid line) and 685 field SNe Ia (gray, 
dashed line) that had good host fits. Host galaxies in the cluster 
sample tend to present less star formation than those in the field 
sample.}
\label{fig:sSFR_hist}
\end{figure}

Another consistency check for our cluster sample selection method
was to compare its host galaxy properties with the ones obtained for
the field sample (except for the redshift, no other host
property was used during our selection). Galaxies inside clusters
are expected to be older, more massive and to have less star formation.
Figs. \ref{fig:age_hist}, \ref{fig:mass_hist} and \ref{fig:sSFR_hist}
show that such expectancy is met, and a Kolmogorov--Smirnov test 
\citep[K--S test, see][]{Mood74mn} indicated 
that the observed differences are significant: the probability that both 
samples were drawn from the same distribution was $1.7\times10^{-3}$ for the mass, 
$8.1\times10^{-4}$ for the hosts age and $8.3\times 10^{-6}$ for the sSFR. A 
comparison between the samples average properties is shown in Table 
\ref{tab:host-average-properties} and also confirms the patterns we expected. 


\begin{table}
\caption{Average host properties for the subset of the cluster and field 
SNIa samples with good host fits (32 and 685 SNe Ia, respectively). 
The age, mass and specific star formation rate are given in Gyr, log of 
solar masses and log of mass fraction per year.}
\label{tab:host-average-properties}
\begin{tabular}{|l|c|c|c|}
\hline
               & Mean age      & Mean mass         & Mean sSFR       \tabularnewline\hline
Cluster Sample & $5.82\pm0.41$ & $10.698\pm0.087$  & $-13.28\pm0.52$ \tabularnewline
Field Sample   & $4.22\pm0.07$ & $10.348\pm0.023$  & $-11.38\pm0.11$ \tabularnewline
Difference     & $1.60\pm0.42$ & $0.35\pm0.09$     & $-1.90\pm0.53$  \tabularnewline
\hline
\end{tabular}
\end{table}

\subsection{Comparing SNIa samples}
\label{sub:comparing-samples}

To assess possible systematic differences between SNe Ia inside
and outside galaxy clusters, we compared the $x_{1}$, $c$ and $HR$
distributions and their cross-correlations in both samples, along
with their assigned $M$, $\alpha$, $\beta$ and $\sigma_{\mathrm{int}}$ 
parameters. The intrinsic scatter $\sigma_{\mathrm{int}}$ was obtained 
separately for each sample by making their Hubble 
residuals reduced $\chi^2$ go to $\sim1$. For the $x_{1}$, 
$c$ and $HR$ distributions we computed their mean, median, 
standard deviation and median absolute deviation (MAD), which is 
defined for a sample $s=\{x_1,x_2,...,x_N\}$ as the median of 
$\{|x_1-x_{\mathrm{med}}|,|x_2-x_{\mathrm{med}}|,...,|x_N-x_{\mathrm{med}}|\}$ 
where $x_{\mathrm{med}}$ is the median of $s$. More attention was given to the
median and MAD during the analysis since they are less sensitive to outliers.

To determine the significance of any difference observed on these properties,
we used a resampling method of selecting from 5,000 to 20,000 random samples with
the same size as the cluster sample from the combination of the field and 
cluster samples, and computing their properties. The fraction $P_{\mathrm{r}}$
of random samples presenting values equal to or more extreme than
the ones obtained for cluster sample (known as $p$-value) yielded 
the probability that the difference observed is due to statistical fluctuations.
In some cases, the SNIa set from which the random samples were drawn 
had a different composition. To avoid possible confusion, these cases are 
explained as they appear and its $P_{\mathrm{r}}$ are marked with special 
superscripts.
 
Differences in cross-correlations were assessed by fitting a linear
model to the two parameters in question and comparing the line slope
for both samples. The fitting was done minimising the $\chi^{2}$; this 
was accomplished after propagating the errors on the x-axis to the y-axis
using a slope obtained from the data assuming equal weights to all
data points. The significance of any slope difference was determined
also by resampling.

We also searched for possible differences in correlations between SNIa
parameters and host-galaxy properties -- mass, age and specific star
formation rate (sSFR) -- using the same method described here. The
errors for the host-galaxy parameters, especially sSFR and mass, are
asymmetrical. In this case, the linear model fitting used an average
of both errors.

\section{Relations between SNe Ia and their host galaxies}
\label{sec:known-relations}

To validate our methods and to have a basis for comparison 
when studying cluster SNIa properties, we first separated the SNe Ia by their 
host's sSFR ($\mathrm{sSFR}<-11.72$ are called `passive' and 
$\mathrm{sSFR}>-10.5$, `active') and checked how reported 
relations between SNIa properties and that of their hosts appeared in 
our data. As mentioned in section \ref{sec:Introduction}, the best established 
reported relations between SNe Ia and their hosts are:

\begin{enumerate}
\item no clear difference in colour distribution was identified 
between SNe Ia in passive and active hosts;
\item SNe Ia in passive hosts have a faster-declining light-curve 
(smaller mean $x_1$);
\item \label{it:alpha} the $\alpha$ parameter is the same regardless of the SNIa host;
\item \label{it:beta} the $\beta$ parameter for passive galaxies is lower 
than that for SNe Ia in active galaxies;
\item \label{it:brighter} SNe Ia in passive galaxies are $\sim0.1$ mag 
more luminous after corrections based on stretch and colour 
(their $M$ in Eq. \ref{eq:distmod_relation} is more negative);
\item when fitted separately, SNe Ia in passive galaxies present less 
scatter on the Hubble residuals than SNe Ia in active galaxies.
\end{enumerate}
When fitted with the same nuisance parameters, item \ref{it:brighter} manifests 
itself through an offset on the Hubble diagram between the SNe Ia in passive and 
active galaxies, and through a correlation between Hubble residuals and host galaxy 
mass or age. 

\begin{figure}
\includegraphics[width=1\columnwidth]{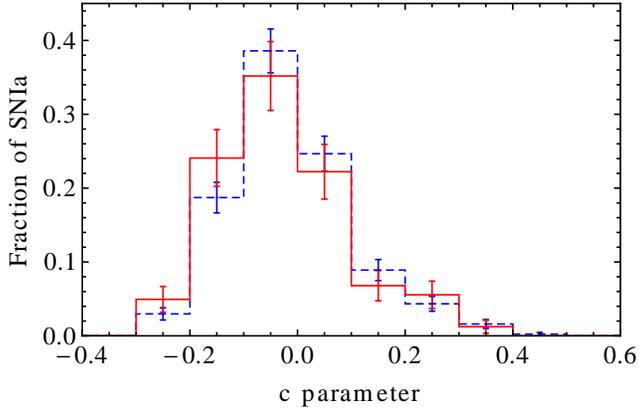}
\caption{Histogram of the colour parameter $c$ distribution for SNe Ia in 
passive (solid red line) and active (blue dashed line) galaxies. Both histograms 
are consistent within the error bars.}
\label{fig:c_passive_active} 
\end{figure}

\begin{figure}
\includegraphics[width=1\columnwidth]{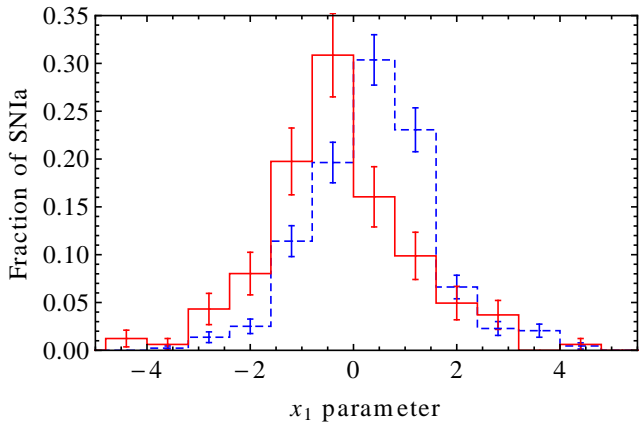}
\caption{Histogram of the stretch parameter $x_1$ distribution for SNe Ia in 
passive (solid red line) and active (blue dashed line) galaxies. The passive 
sample histogram is clearly shifted towards lower values of $x_1$. It also presents 
a larger scatter, probably due to its tails.}
\label{fig:x1_passive_active} 
\end{figure}

Figs. \ref{fig:c_passive_active} and \ref{fig:x1_passive_active} show that while $c$ distributions 
for SNe Ia in passive and active galaxies do not present any significant differences, $x_1$ 
distributions are clearly different. A K--S test indicated that the probability that both $c$ 
samples are drawn from the same distribution is 0.47, while such probability is $3.4\times 10^{-12}$ 
for $x_1$. Table \ref{tab:x1-c-passive-active}, however, presents some tension between the 
samples average colours, although this is not as significant as the difference in the average $x_1$. 
It is also possible to notice that the 
$x_1$ distribution in passive galaxies is significantly broader, a rarely reported result. 
It is in qualitative agreement with \citet[fig. 6]{Smith12mn}, but in qualitative 
disagreement with \citet[fig. 12]{Sullivan06mn}.


\begin{table}
\caption{Statistical measures for the $x_1$ and $c$ distributions 
of 162 and 438 SNe Ia in passive and active hosts. $P_{\mathrm{r}}^{\mathrm{*}}$ 
is the fraction of 20,000 random samples of SNe Ia in active galaxies, of 
same size as the passive sample, that present a value more extreme than the one 
observed for the passive sample.}
\label{tab:x1-c-passive-active}
\begin{tabular}{|c|c|c|c|c|}
\hline
$x_1$   & Mean                & Median              & Std. Dev. & MAD   \tabularnewline\hline
Passive &              -0.328 & -0.475              & 1.41      & 0.847 \tabularnewline 
Active  &               0.382 &  0.381              & 1.18      & 0.695 \tabularnewline
$P_{\mathrm{r}}^{\mathrm{*}}$ & $\la 5\times 10^{-5}$ & $\la 5\times 10^{-5}$ & $\la 5\times 10^{-5}$ & 0.001 \tabularnewline
\hline
$c$           & Mean    & Median  & Std. Dev. & MAD   \tabularnewline\hline
Passive       & -0.0278 & -0.0439 & 0.120     & 0.070 \tabularnewline
Active        & -0.0135 & -0.0294 & 0.117     & 0.066 \tabularnewline
$P_{\mathrm{r}}^{\mathrm{*}}$ &  0.027  &  0.024  & 0.320     & 0.217 \tabularnewline
\hline
\end{tabular}
\end{table}

\begin{figure}
\includegraphics[width=1\columnwidth]{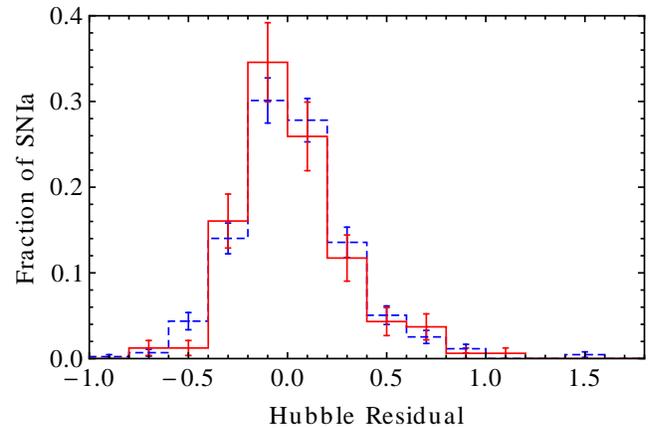}
\caption{Histogram of the Hubble residuals for SNe Ia in passive (solid red line) 
and active (blue dashed line) galaxies obtained with each samples particular 
nuisance parameters. No clear difference can be noted.}
\label{fig:hr_passive_active} 
\end{figure}

When fitting for SNe Ia nuisance parameters separately in the passive and active sample,
we obtained compatible values of $\alpha$ but significantly different values for $\beta$ 
and the average absolute magnitude $M$. Table \ref{tab:nuisance-passive-active} shows 
that our results are compatible with previously reported ones -- 
items \ref{it:alpha} to \ref{it:brighter} above. The 
difference in Hubble residuals scatter, however, could not be detected. The MAD calculated for 
the active and passive sample were 0.167 and 0.152, with a probability 
$P_{\mathrm{r}}^{\mathrm{*}}$ for the active sample to reach the passive sample deviation of 
0.241. A histogram of this distribution is shown in Fig. \ref{fig:hr_passive_active}. 

\begin{table}
\caption{Nuisance parameters, the Hubble residuals median absolute deviation $D_{\mathrm{HR}}$ 
and the intrinsic scatter obtained for the 162 and 438 SNe Ia in passive and active galaxies. 
$P_{\mathrm{r}}^{\mathrm{*}}$ was calculated from 5,000 random samples drawn 
from the active sample.}
\label{tab:nuisance-passive-active}
\begin{tabular}{|c|c|c|c|c|c|}
\hline
                        & $\alpha$  & $\beta$       & $M$                 &$D_{\mathrm{HR}}$  & $\sigma_{\mathrm{int}}$\tabularnewline\hline
Active                  & 0.180(18) & 3.35(14)      & -19.305(13)         & 0.167  & 0.17                \tabularnewline  
Passive                 & 0.206(20) & 2.54(22)      & -19.420(22)         & 0.152  & 0.16                \tabularnewline  
$P_{\mathrm{r}}^{\mathrm{*}}$& 0.229     &$2\times10^{-4}$& $\la 2\times10^{-4}$ & 0.241  & 0.396               \tabularnewline\hline
\end{tabular}
\end{table}

\begin{figure}
\includegraphics[width=1\columnwidth]{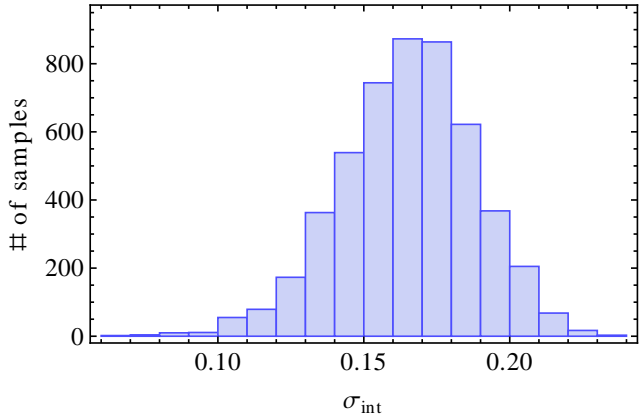}
\caption{Histogram of the intrinsic scatter obtained for 5,000 randomly selected 
samples of 162 SNe Ia residing in active galaxies. For each random sample, 
$\sigma_{\mathrm{int}}$ was obtained by setting the minimum reduced chi-squared 
to $\sim 1$.}
\label{fig:active_sigint} 
\end{figure}

The difference in the intrinsic scatter was not significant either, as Table 
\ref{tab:nuisance-passive-active} shows. This can also be noticed from Fig. 
\ref{fig:active_sigint}, which presents the $\sigma_{\mathrm{int}}$ values obtained 
for the 5,000 randomly drawn samples of 162 
SNe Ia hosted by active galaxies (the `active sample'). Given the large 
variability found for $\sigma_{\mathrm{int}}$ in these samples, even if the passive 
sample had a smaller value like $\sigma_{\mathrm{int}}=0.13$ (which could seem 
significant), it could still be achieved by 6.7 per cent of our active samples 
of same size. 

The lack of detectable difference in the Hubble residuals scatter could be due 
to the exclusion of outliers through the 4$\sigma$ cut and to the use of the 
median absolute deviation instead of a standard deviation. In fact, when including 
the 10 outliers beyond 4$\sigma$ and comparing standard deviations, our active sample 
showed significantly larger scatter than the passive sample. However, we 
disregarded this result as it may be caused by non--Ia contamination. 

We compared our passive and active nuisance parameter values and intrinsic scatter 
with those obtained by \citet{Lampeitl10mn} using the same resampling methodology. 
For instance, we selected, from our passive sample, 5,000 random samples containing 
40 SNe Ia (same size as their passive sample) and counted the fraction that could 
reach their values. As Table \ref{tab:lampeitl-comparison} shows, all parameters 
were compatible for both samples. 

\begin{table}
\caption{Nuisance parameters and intrinsic scatter obtained by \citet{Lampeitl10mn} 
for a sample of 40 SNe Ia in passive hosts and 122 SNe Ia in active hosts, and 
the probability that our data could reach these values. $P_{\mathrm{r}}^{\mathrm{*}}$ 
was obtained from 5,000 random samples drawn from our active sample and 
$P_{\mathrm{r}}^{\mathrm{-}}$ was obtained from 5,000 random samples drawn from 
our passive sample. Their average magnitudes were corrected for the difference 
in the assumed $H_0$ and for the 10.635 offset between {\sc SALT2mu} output and $B$ band 
AB magnitude.}
\label{tab:lampeitl-comparison}
\begin{tabular}{|r|c|c|c|c|c|}
\hline
                         & $\alpha$  & $\beta$  & $M$        & $\sigma_{\mathrm{int}}$\tabularnewline\hline
Lampeitl's Active        & 0.12(01)  & 3.09(10) & -19.30(01) & 0.17                \tabularnewline  
$P_{\mathrm{r}}^{\mathrm{*}}$ & 0.071     & 0.162    & 0.480      & 0.258               \tabularnewline\hline
Lampeitl's Passive       & 0.16(02)  & 2.42(16) & -19.39(03) & 0.13                \tabularnewline
$P_{\mathrm{r}}^{\mathrm{-}}$ & 0.119     & 0.340    & 0.286      & 0.483               \tabularnewline\hline  
\end{tabular}
\end{table}

\section{Cluster SNe Ia properties}
\label{sec:cluster-sn-properties}

\subsection{Comparison with field sample}
\label{sub:comp-field-sample}

\begin{figure}
\includegraphics[width=1\columnwidth]{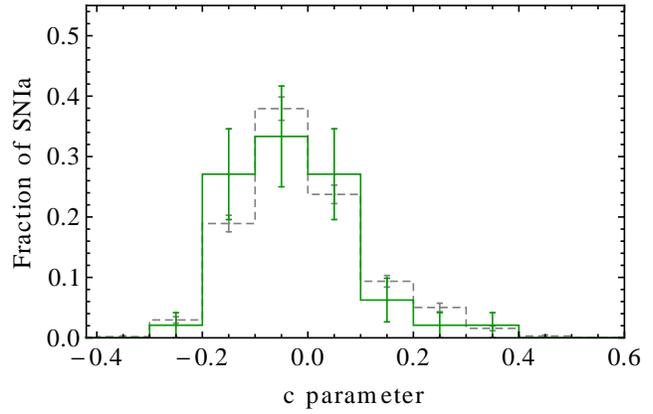}
\caption{Histogram of the SALT2 colour parameter for SNe Ia inside 
galaxy clusters (solid green line) and outside (gray dashed line). No 
significant difference is seen.}
\label{fig:c-cluster-field} 
\end{figure}

\begin{figure}
\includegraphics[width=1\columnwidth]{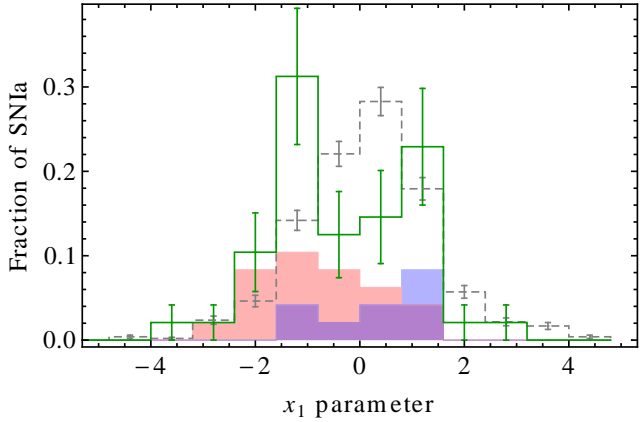}
\caption{Histogram of the SALT2 stretch parameter for SNe Ia inside 
galaxy clusters (solid green line) and outside (gray dashed line). 
The difference between the distributions is significant and our cluster 
sample is bimodal in $x_1$. The red and blue shaded histograms present the 
cluster SNe Ia that had hosts typed as passive and active, 
respectively (their overlap is depicted in purple). 
Each shaded histogram is associated with a different cluster sample peak.}
\label{fig:x1-cluster-field} 
\end{figure}

When comparing SNe Ia inside and outside rich galaxy clusters, no significant 
difference was found in the colour distribution (see Fig. \ref{fig:c-cluster-field} 
and Table \ref{tab:x1-c-cluster-field}). A K--S test, which returned a $p$-value of 0.74, 
did not indicate any differences either. However, there are several indications that 
the $x_1$ distribution in cluster SNe Ia differs from that of the field SNe Ia. A probability 
of 0.0021 was obtained from a K--S test with the null hypothesis that the two samples were 
drawn from the same distribution, and Table \ref{tab:x1-c-cluster-field} shows 
that the cluster $x_1$ distribution is shifted to lower values. Furthermore, 
Fig. \ref{fig:x1-cluster-field} suggests that the cluster sample $x_1$ 
distribution is bimodal, with the left peak consisting of 
mostly SNe Ia in passive hosts and the right peak consisting 
of mostly SNe Ia in active hosts. At least part of the SNe Ia in the right peak 
are actually contamination from the field, which has a high concentration of active 
galaxies ($\sim63$ per cent in our sample).

\begin{figure}
\includegraphics[width=1\columnwidth]{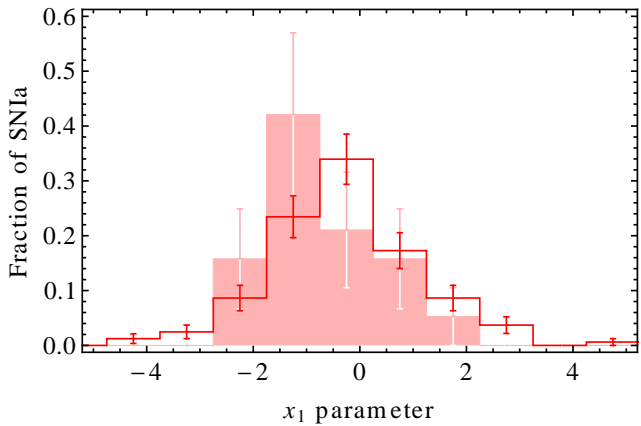}
\caption{Comparison between $x_1$ distributions of the full passive sample 
(red contour, no filling) and the sub-sample of SNe Ia in passive galaxies inside 
rich galaxy clusters (no contour, red filling). The peak of the cluster histogram 
is shifted towards lower values of $x_1$.}
\label{fig:x1-passive-in-out} 
\end{figure}

The position of the left peak in Fig. \ref{fig:x1-cluster-field}, however, 
does not coincide with the position obtained for the passive sample depicted 
in Fig. \ref{fig:x1_passive_active}: it is slightly more negative. 
To assess the significance of this difference, we compared the statistical 
properties of SNe Ia in passive hosts selected as being inside clusters 
with the full passive sample (see Fig. \ref{fig:x1-passive-in-out}). The 
mean and median $x_1$ obtained for the former sample were -0.82 and -1.12, and 
the fraction of random samples of same size (19 SNe Ia) drawn from the full 
passive sample which could reach lower values were 0.05 and 0.006, respectively. 
This result indicates that the cluster environment may intensify the passive bias towards 
fast-declining SNe Ia, possibly by preferentially selecting very old hosts. This is 
investigated in section \ref{sec:host-age}. A similar comparison was performed 
for the $x_1$ distribution of SNe Ia in active galaxies, but no significant difference was 
found.

\begin{table}
\caption{Statistical measures for the $x_1$ and $c$ distributions 
of the 48 SNe Ia inside and 1015 outside clusters. $P_{\mathrm{r}}$ 
was obtained from 20,000 random samples.}
\label{tab:x1-c-cluster-field}
\begin{tabular}{|c|c|c|c|c|}
\hline
$x_1$            & Mean               & Median            & Std. Dev. & MAD   \tabularnewline\hline
Field            & 0.142              & 0.202             & 1.27      & 0.763 \tabularnewline 
Cluster          & -0.398             & -0.602            & 1.38      & 0.940 \tabularnewline
$P_{\mathrm{r}}$   &  $1.9\times10^{-3}$ &  $2\times 10^{-4}$ & 0.232     & 0.104 \tabularnewline
\hline
$c$             & Mean    & Median  & Std. Dev. & MAD   \tabularnewline\hline
Field           & -0.0113 & -0.0270 & 0.121     & 0.069 \tabularnewline
Cluster         & -0.0251 & -0.0360 & 0.113     & 0.073 \tabularnewline
$P_{\mathrm{r}}$   &  0.218  &  0.301  & 0.311     & 0.332 \tabularnewline
\hline
\end{tabular}
\end{table}

For the nuisance parameters, Hubble residuals and intrinsic scatter, no 
significant difference could be 
identified between the cluster and the field sample, although $\beta$, $M$, 
$\sigma_{\mathrm{int}}$ and the Hubble residuals scatter follow the same trend as the 
full passive sample relative to the full active sample. 
As Table \ref{tab:nuisance-cluster-field} shows, the probability 
$P_{\mathrm{r}}$ was fairly high for every parameter. 

\begin{table}
\caption{Nuisance parameters, Hubble residuals median absolute deviation $D_{\mathrm{HR}}$ 
and the intrinsic scatter obtained for the cluster and field samples (containing 48 
and 1070 SNe Ia, respectively). The quantity $P_{\mathrm{r}}$ was obtained from 5,000 
random samples.}
\label{tab:nuisance-cluster-field}
\begin{tabular}{|c|c|c|c|c|c|}
\hline
                & $\alpha$  & $\beta$  & $M$         & $D_{\mathrm{HR}}$ & $\sigma_{\mathrm{int}}$ \tabularnewline\hline
Field           & 0.180(09) & 3.26(08) & -19.337(08) & 0.155  &  0.16                \tabularnewline  
Clusters        & 0.156(22) & 2.46(32) & -19.389(30) & 0.115  &  0.13                \tabularnewline  
$P_{\mathrm{r}}$   & 0.320     & 0.085    & 0.129       & 0.104  &  0.556               \tabularnewline
\hline
\end{tabular}
\end{table}

Due to the small size of the sample (see Table \ref{tab:summary-numbers}), we did not perform 
a nuisance parameters fit for the sub-sample of SNe Ia in passive hosts in rich clusters. 
Such fits involve many parameters and thus the uncertainties 
would be quite large. We instead fitted for the whole passive sample and compared 
the mean values of the Hubble residuals for the SNe Ia belonging to clusters and to the field. 
Although the cluster sub-sample shows an offset 
from the field sub-sample of $-0.084$, 10 per cent of randomly selected samples of 
SNe Ia in passive hosts were able to mimic such difference. A larger cluster sample 
is necessary to determine if such offset is significant.

\subsection{Comparison with mixed samples}
\label{sec:mixed-samples}

Given that SNe Ia in passive and active hosts are known to differ, it is 
important to test if the cluster sample properties can be mimicked by 
samples with the same fraction of passive and active hosts, which we
call `mixed samples'. Since 4.2 per cent of all SNe Ia did not have an 
identified host and 30.4 per cent of the remaining ones had hosts with 
bad fits, the composition of the cluster sample is not known precisely. Based 
on the fitted hosts, we estimated that the cluster sample should be composed 
of approximately 65 per cent passive and 35 per cent active hosts.

\begin{figure}
\includegraphics[width=1\columnwidth]{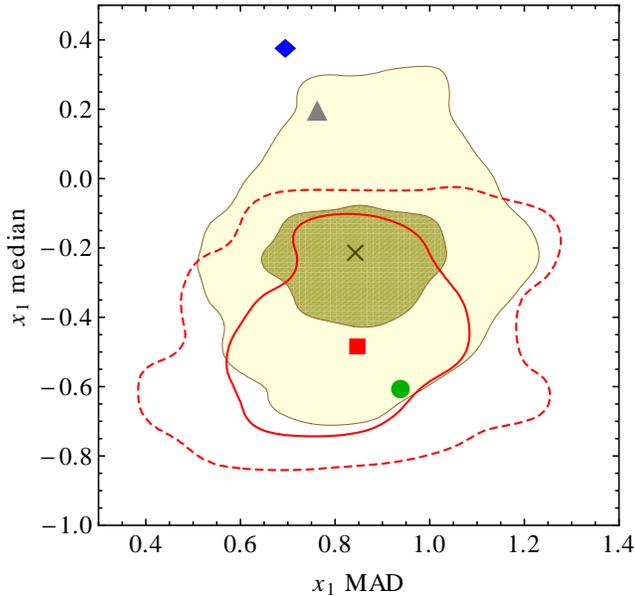}
\caption{Comparison between samples $x_1$ median and MAD. The blue diamond and red square 
represent the active and passive samples, and the gray triangle and the green circle 
represent the field and cluster samples, respectively. The shaded regions represent 
the parameter space populated by 68 (dark shade) and 95 (light shade) per cent of the 
randomly selected mixed samples containing 70 per cent of passive hosts, and the cross 
indicate their mean value. The red contours 
represent the region populated by 68 (solid line) and 95 per cent (dashed line) of randomly 
selected samples from the passive sample. All random samples contain 48 SNe Ia. The cluster 
sample is barely consistent with samples with 70 per cent of passive hosts. A higher passive 
fraction is preferred.}
\label{fig:confidence-regions} 
\end{figure}

Comparisons were made to samples with a range of compositions, from 50 to 100 
per cent of passive hosts. In all comparisons, the colour distribution and 
the nuisance parameters from both samples were compatible, 
while the $x_1$ distribution for the cluster sample had 
a mean and median value lower than every mixed sample. Since a higher fraction 
of passive hosts shifts the $x_1$ distribution to lower values, the compatibility 
of the $x_1$ distributions continually increases with the fraction of 
passive hosts in the mixed sample. As Table \ref{tab:mixed-samples} and Fig. 
\ref{fig:confidence-regions} show, it is unlikely that a mixed sample with a 
similar composition as the cluster sample could reach such a low value for 
the $x_1$ median. Ignoring the composition estimate for the cluster sample, it 
would be most compatible with the pure passive sample. An intermediate sample 
composition would probably be a better choice to explain the cluster sample. 
However, in section \ref{sec:cluster-radius} we show that for SNe Ia closer to the 
cluster's centre the difference in $x_1$ is significant even for samples with 
high passive host content.

\begin{table}
\caption{Average values obtained from 5,000 mixed samples composed by 
70 per cent of passive and 30 per cent of active hosts (leftmost columns) 
and from 5,000 mixed samples composed by 100 per cent of passive hosts 
(rightmost columns). $P_{\mathrm{r}}^{70}$ and $P_{\mathrm{r}}^{-}$ give their 
respective fractions that could reach the cluster sample's values.}
\label{tab:mixed-samples}
\begin{tabular}{|l|c|c|l|c|c|}
\hline
0.7 passive          & Value  & $P_{\mathrm{r}}^{70}$ & 1.0 passive          & Value  & $P_{\mathrm{r}}^{-}$ \tabularnewline\hline
$\alpha$             & 0.194  & 0.193 & $\alpha$             & 0.221  & 0.072 \tabularnewline
$\beta$              & 2.88   & 0.224 & $\beta$              &  2.64  & 0.355 \tabularnewline
$M$                  & -19.37 & 0.386 & $M$                  & -19.43 & 0.215 \tabularnewline
HR MAD               & 0.16   & 0.049 & HR MAD               & 0.16   & 0.038 \tabularnewline
$\sigma_{\mathrm{int}}$ & 0.13   & 0.446 & $\sigma_{\mathrm{int}}$ & 0.12   & 0.488 \tabularnewline
$c$ median           & -0.040 & 0.393 & $c$ median           & -0.042 & 0.426 \tabularnewline
$c$ MAD              &  0.067 & 0.279 & $c$ MAD              & 0.069  & 0.344 \tabularnewline
$x_1$ median         & -0.179 & 0.019 & $x_1$ median         & -0.458 & 0.241 \tabularnewline
$x_1$ MAD            &  0.86  & 0.241 & $x_1$ MAD            & 0.80   & 0.129 \tabularnewline
\hline
\end{tabular}
\end{table}

\subsection{The role of the host age}
\label{sec:host-age}

\begin{figure}
\includegraphics[width=1\columnwidth]{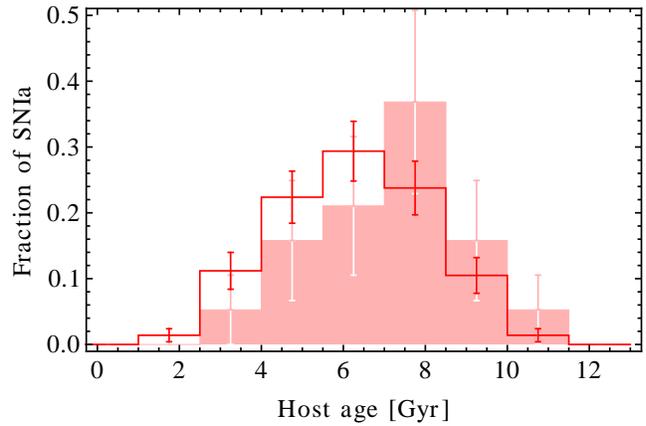}
\caption{Comparison between host age distributions of the full passive sample 
(red contour, no filling) and the sub-sample of SNe Ia in passive galaxies inside 
rich galaxy clusters (no contour, red filling). The distribution for the cluster 
sample is shifted to larger values, although its significance is low.}
\label{fig:age-passive-in-out} 
\end{figure}

The relationship between galaxy age and environment density is a 
well established fact: passive galaxies in high-density regions -- 
such as rich clusters -- are, on average, $\sim2$ Gyr older than 
passive galaxies in low-density regions, as shown by 
\citet{Thomas05mn} using spectra of 54 early-type galaxies in 
high-density and 70 in low-density environments. Based only in our 
photometric data, we searched for age differences between passive 
galaxies inside and outside rich clusters, and Figure \ref{fig:age-passive-in-out} 
shows that the cluster sample passive hosts were estimated to be, 
on average, older than the field passive hosts. However, the significance of 
such result is low since a K--S test returned a $p$-value of 0.157. Moreover, 
by resampling 20,000 times from the passive sample, we obtained 1206
(6 per cent) samples that could reach an average age higher than the 
cluster sample. The lack of significance in this detection is probably 
due to larger errors in the classification of galaxies and in the 
determination of their ages.

\begin{figure}
\includegraphics[width=1\columnwidth]{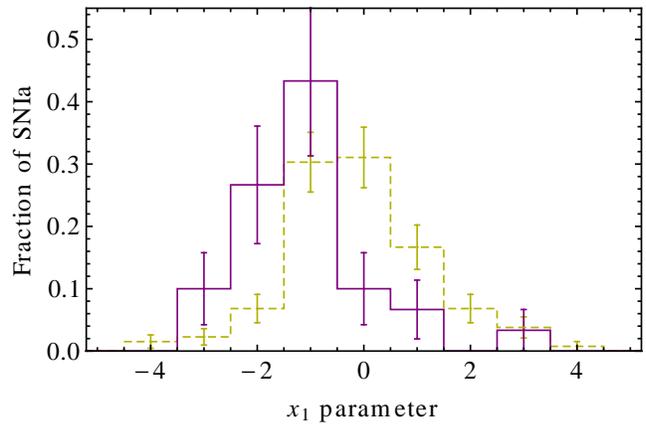}
\caption{Comparison between $x_1$ distributions of SNe Ia hosted by old ($>8$ Gyr) passive 
galaxies (solid purple line) and of SNe Ia hosted by young ($<8$ Gyr) passive galaxies 
(dashed yellow line). SNe Ia in old passive galaxies have, on average, a smaller stretch 
than those in young passive galaxies.}
\label{fig:x1-age-hist} 
\end{figure}

The indication that passive galaxies inside rich clusters may host
SNe Ia with smaller stretch than passive galaxies outside clusters
prompted us to study possible causes for such difference. First of
all, significant trends between SNIa stretch and host age have been
reported by \citet{Gupta11mn}, that showed that older galaxies
host SNe Ia with smaller $x_1$. 
While this result may be attributed to the differences presented in section 
\ref{sec:known-relations} (since older galaxies are usually passive), such trend 
seems to remain within passive galaxies: \citet{Gallagher08mn} have 
pointed out that early-type galaxies older than 5 Gyr host SNe Ia that are 
$\sim 1$ mag fainter than those in younger early-type galaxies. This result 
means that the stretch of SNe Ia in old passive galaxies should populate 
lower values than those in young passive galaxies. This conclusion was 
confirmed by our passive sample, as Fig. \ref{fig:x1-age-hist} shows. 
While the difference is still significant for the age cut of 5 Gyr proposed 
by \citet{Gallagher08mn}, our results are stronger using a separation at 
8 Gyr, which created samples containing 132 (host age $<8$ Gyr, called `young') and 
30 (host age $>8$ Gyr, called `old') SNe Ia. 
A K--S test between SNe Ia $x_1$ distribution in young and old passive galaxies 
returned a $p$--value of $3.4\times10^{-5}$.

\subsection{Correlation with host galaxy mass}

\begin{figure}
\includegraphics[width=1\columnwidth]{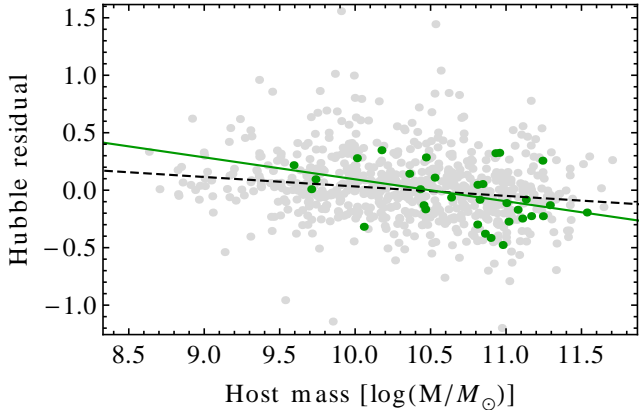}
\caption{Correlation between SNIa Hubble residuals and its host mass for 
the field (gray dots) and cluster (green dots) samples. The black dashed and green solid 
lines are linear fits adjusted to the field and cluster samples, with 
slopes $-0.082\pm 0.015$ and $-0.191\pm 0.078$, respectively. Both samples show 
anti-correlations that are compatible at 7.5 per cent. The errors bars were hidden to 
facilitate visualisation.}
\label{fig:correlation-hr-mass} 
\end{figure}

Many authors have shown the correlation between SNIa Hubble residuals and 
their host galaxy's mass \citep[e.g.][]{Gupta11mn, Lampeitl10mn}. Fig. \ref{fig:correlation-hr-mass} 
presents the correlation obtained for our sample of field and cluster SNe Ia; 
both groups present this same trend and they are compatible, since a probability $P_{\mathrm{r}}$ for 
the cluster sample slope was calculated at 7.5 per cent, not a high significance level.

\section{Robustness tests}
\label{sec:rubustness-tests}

To verify how our conclusions depend on our methods, we tested how the 
use of stringent cuts on $x_1$ and $c$ parameters, the cluster sample redshift 
distribution, the assumed cluster radius and the host galaxy photometry 
used could affect our findings.

\subsection{$x_1$ and $c$ cuts}
\label{sub:x1-c-cuts}

\begin{figure}
\includegraphics[width=1\columnwidth]{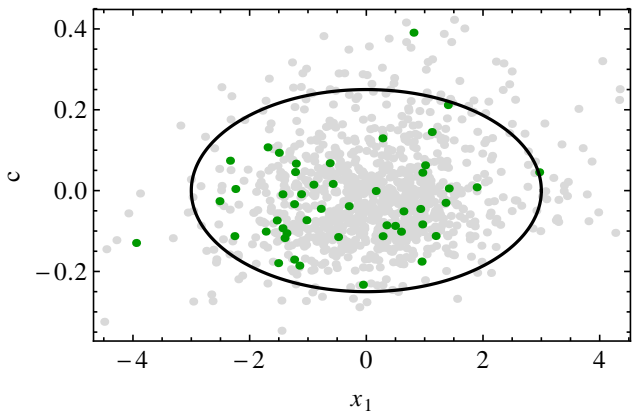}
\caption{Distribution of SNe Ia on the $x_1$--$c$ parameter space and the region 
chosen to be analysed separately (black line). The ellipse is centred in 
($x_1=0,c=0$) and has semi-major axis $a_{x_1}=3$ and semi-minor axis $a_c=0.25$. 
The gray and green dots represent field and cluster SNe Ia, respectively.}
\label{fig:x1-c-cuts}
\end{figure}

To test whether our results depend on the core or on the tails of SNe Ia 
distributions in $x_1$ and $c$ and to further remove contamination by 
non--Ia SNe, we repeated our analysis after applying the elliptical 
cut suggested by \citet{Campbell13mn}. This cut -- presented in Fig. 
\ref{fig:x1-c-cuts} -- was chosen through simulations and is expected 
to remove a larger fraction of non--Ia than of Ia SNe. Table 
\ref{tab:x1-c-cut-numbers} presents the sample sizes after this cut.

\begin{table}
\caption{Amount of SNe Ia in the cluster and field samples after selecting 
those inside the ellipse depicted in Fig. \ref{fig:x1-c-cuts}.}
\label{tab:x1-c-cut-numbers}
\begin{tabular}{|l|c|c|c|c|}
\hline
         & SNe Ia & w/ host fit & Active & Passive \tabularnewline\hline
Cluster  & 45     & 30          & 8      & 19      \tabularnewline
Field    & 896    & 602         & 385    & 122     \tabularnewline
\hline
\end{tabular}
\end{table}

This cut did not affect any of our conclusions since the significance of the 
parameters difference did not change much. One more subtle result was the 
comparison between the passive and active samples $x_1$ distribution width. 
After the cut, this difference was only noticeable in the distribution's 
standard deviation, as $P_{\mathrm{r}}^*$ for the MAD got to 0.199. 
This result corroborates the hypothesis that the passive sample include 
some extra SNe Ia on its $x_1$ distribution's tails, but which are still 
present in the $-3\la x_1\la 3$ range.  

When comparing the full passive sample with its sub-sample residing in 
rich clusters, the significance of the difference between the $x_1$ mean 
and median values was slightly strengthened: the fractions $P_{\mathrm{r}}^{-}$ 
of random samples drawn from SNe Ia in passive galaxies that could surpass the 
cluster sub-samples $x_1$ mean and median were 0.011 and 0.002, indicating that 
such difference is not caused by non--Ia contamination. All other $x_1$ 
comparisons maintained similar significance levels.

\subsection{Redshift dependence}

Even though the redshift distribution of the cluster and field samples are not very different, 
we checked if our results could be caused by redshift selection. For that we counted the number 
of cluster SNe Ia in 0.05 wide redshift bins and randomly selected the same number of SNe Ia 
in each bin from our full sample, creating 5,000 random samples with same size and redshift 
distribution as the cluster sample, which we called `same--z samples'. 
Our significance analysis was then repeated using these samples.

All the results from section \ref{sub:comp-field-sample} were reobtained at very similar 
significance levels. We did not perform this test for the comparison between the cluster sample 
and the mixed samples since the passive sample is not big enough to adequately sample the 
parameter space after being broken into bins of redshift. 

As a consistency check, we also looked for possible differences between the same--z samples and the 
original full sample. Not a single property was found to differ significantly, as for 
all of them, at least 10 per cent of the same--z samples were able to achieve the full sample 
values. This test shows that the effect of redshift selection is unlikely to cause the 
differences we observed.

\subsection{Only optical photometry}

Since combining host galaxy photometry from different surveys might cause a few 
problems like matching wrong objects across the catalogues and small differences 
in the magnitude aperture and calibration, we also performed our analysis using 
host properties obtained from the SDSS photometry alone. The effects of 
excluding \emph{GALEX} and UKIDSS from the galaxy properties determination was 
investigated in \citet{Gupta11mn} and in general it increases the errors for 
all host parameters, specially for sSFR. In addition, it compresses the estimated 
ages to a smaller range centred around 6 Gyr. Such effects were also verified in this work. 
The elimination of \emph{GALEX} and UKIDSS photometry can only affect analysis 
involving host galaxy properties, therefore all conclusions regarding the differences 
between the field and cluster samples remains unaltered.

While the significance levels of our results changed when using SDSS photometry only, 
these changes were small and, for most cases, did not affect our conclusions. All 
differences between the active and passive sample remained significant, as well as 
the difference in the $x_1$ distribution for SNe Ia in old and young passive galaxies. 
The difference in $x_1$ median between the cluster (using the maximum separation 
of 1.5 Mpc) and the mixed samples, however, lost its significance since the 
fraction of mixed samples with 70 per cent passive hosts that 
could surpass the cluster sample $x_1$ median was 0.07. This result is, therefore, 
dependent on the use of \emph{GALEX} and UKIDSS photometry. When comparing the sample 
closer to the cluster core (within 1.0 Mpc from the centre -- see section 
\ref{sec:cluster-radius}) to the mixed samples, the significance of this difference 
remained. 

\subsection{Cluster radius}
\label{sec:cluster-radius}

We repeated our analysis using a physical radius of 1.0 Mpc during the projection 
test described in section \ref{sub:projection-test}, which reduced the cluster sample 
to 31 SNe Ia. This change can only affect analysis involving the cluster sample, 
thus comparisons between the active and passive sample and between SNe Ia inside 
old and young passive galaxies were not affected. Most of the results related to the 
cluster sample were consistent with the ones obtained using 1.5 Mpc. Few exceptions 
appeared in the form of an intensification of previous signals, which may indicate that 
some contamination from the field can be eliminated with a smaller radius, or that 
SNe Ia closer to the core of clusters are the ones responsible for the differences 
seen between the field and cluster samples. It is important to note that, with the 
reduction of the radius, the estimated fraction of passive galaxies increase to 
approximately 75 per cent. This corroborates our conclusion that the SNe Ia responsible 
for the differences observed reside in passive galaxies. Moreover, the strengthening of 
previous signals presented in this section qualitatively agrees with the connection 
between host age and SNe Ia properties presented in section \ref{sec:host-age}, since 
older galaxies are expected to populate the inner, denser regions of clusters 
\citep{Balogh99mn,Thomas05mn}.

The cluster sample $x_1$ distribution got shifted to smaller values, increasing the 
difference when compared to the field and mixed samples. Its mean and median got to 
-0.75 and -1.12, respectively, and the probability $P_{\mathrm{r}}$ for getting this same result 
from the full SNe Ia sample was $10^{-4}$ for the mean and less than that for the median. 
When comparing with the full passive sample, $P_{\mathrm{r}}$ was calculated as 0.026 for the $x_1$ 
mean and $6\times10^{-4}$ for the $x_1$ median. Table \ref{tab:r1mpc} compares the values 
obtained for the 1.5 and the 1.0 Mpc cluster radius and their probabilities $P_{\mathrm{r}}$ and 
$P_{\mathrm{r}}^-$ for drawing these values from the full and the passive samples, respectively.

\begin{table}
\caption{Mean and median SALT2 parameters $x_1$ and $c$ obtained for the field sample, 
the passive sample, and the cluster sample using 1.5 and 1.0 Mpc physical radius during 
the projection test described in section \ref{sub:projection-test}. Following each cluster 
sample values, we present the probabilities $P_{\mathrm{r}}$ and $P_{\mathrm{r}}^-$ that a more 
extreme value could be reached by random selections from the full SNe Ia sample and the 
passive sample.}
\label{tab:r1mpc}
\begin{tabular}{|l|c|c|c|c|c|}
\hline
                & $x_1$ mean         & $x_1$ median      & $c$ mean & $c$ median \tabularnewline\hline
Field           & 0.142              & 0.202             & -0.0113  & -0.0270    \tabularnewline
Passive         & -0.328             & -0.475            & -0.0278  & -0.0439    \tabularnewline\hline
1.5 Mpc         & -0.398             & -0.602            & -0.0251  & -0.0360    \tabularnewline 
$P_{\mathrm{r}}$   & $1.9\times 10^{-3}$ & $2\times 10^{-4}$ & 0.218    &  0.301      \tabularnewline
$P_{\mathrm{r}}^-$ & 0.347              & 0.241             & 0.426    & 0.347       \tabularnewline\hline 
1.0 Mpc         & -0.750             & -1.118            & -0.0410  & -0.0454     \tabularnewline   
$P_{\mathrm{r}}$   & $10^{-4}$           & $\la 10^{-4}$      & 0.081   & 0.224       \tabularnewline
$P_{\mathrm{r}}^-$ & 0.026              & $6\times 10^{-4}$  & 0.246   & 0.431        \tabularnewline\hline      

\end{tabular}
\end{table}

\section{Conclusions and summary}
\label{sec:conclusions}

We used SDSS photometrically and spectroscopically typed SNe Ia, 
galaxy photometry from SDSS, \emph{GALEX} and UKIDSS, and the GMBCG optical 
cluster catalogue to study the properties of SNe Ia residing in rich galaxy 
clusters. Their light curves were parametrized by the SALT2 model. 

To test our samples and methods, we first analysed the properties of SNe Ia residing 
in active and passive galaxy hosts and compared our results to the literature (see 
section \ref{sec:known-relations}). We confirm previously reported 
differences between these SNe Ia, namely that passive 
galaxies host SNe Ia that: (a) have smaller average $x_1$ (see Fig. 
\ref{fig:x1_passive_active}); (b) are $\sim 0.1$ mag brighter after 
corrections based on $x_1$ and $c$; and (c) have a smaller SALT2 $\beta$ parameter. 
Moreover, (d) the consistency between SALT2 $\alpha$ parameters obtained for SNe Ia 
in active and passive galaxies was also confirmed. 
Contrary to previous works, we (e) could not detect a significant difference 
in the Hubble residual scatter: even though the Hubble residual intrinsic scatter 
and median absolute deviation were smaller in passive galaxies, they were considered 
consistent with those obtained for SNe Ia in active galaxies. This indicates that 
any differences that might be observed are due to a few objects in the active sample 
that present high dispersion. 
Items (b) through (e) are summarised in Table \ref{tab:nuisance-passive-active}. 
We also report that the $x_1$ distribution of our passive sample is 
significantly broader than that of our active sample (see Table 
\ref{tab:x1-c-passive-active}).

We then analysed SNe Ia residing in rich galaxy clusters, and concluded 
that their SALT2 $x_1$ distribution is different from that of their 
counterparts in the field (see Fig. \ref{fig:x1-cluster-field} and Table 
\ref{tab:x1-c-cluster-field}): it is shifted to lower values, and the 
probability that a lower median could be obtained from a randomly 
selected SNe Ia sample was estimated as $2\times 10^{-4}$. Although this 
could be explained by a higher content of passive galaxies SNe Ia 
than the estimated one (of 65 per cent, see section \ref{sec:mixed-samples}), 
this explanation is not sufficient for regions closer to the core of the cluster 
(see section \ref{sec:cluster-radius}). Moreover, we found evidence that 
passive galaxies inside rich clusters may host SNe Ia with smaller $x_1$ 
than passive galaxies outside them (see Fig. \ref{fig:x1-passive-in-out}). 
This difference could be due to a higher content of old passive galaxies 
in clusters as shown by \citet{Thomas05mn}. As demonstrated in section 
\ref{sec:host-age}, old passive galaxies host SNe Ia with smaller $x_1$ than young 
passive galaxies (see Fig. \ref{fig:x1-age-hist}), a result compatible with 
the findings by \citet{Gallagher08mn}. 

Other cluster SNe Ia parameters -- their colour distribution, nuisance parameters 
and Hubble residual scatter -- were found to be consistent with those of field 
SNe Ia, although they all followed the same trends as SNe Ia in passive galaxies 
do when compared to SNe Ia in active galaxies. To verify if differences exist in 
these parameters, larger samples are required. Current and near-future projects 
like DES\footnote{\texttt{http://www.darkenergysurvey.org}}, 
Pan-STARRS\footnote{\texttt{http://pan-starrs.ifa.hawaii.edu}} and 
J-PAS\footnote{\texttt{http://j-pas.org/}} 
are expected to increase the SNIa sample sizes by a factor of 5 or more. 
The combination of the SDSS dataset with other datasets, 
provided these overlap with a cluster catalogue, would also increase 
sample sizes. However, this combination might encounter difficulties  
due to possible systematic differences between the datasets. An interesting 
possibility to better constrain these results would be to assess host properties 
using spectroscopy, which exists for all SNe Ia used here, thanks to the BOSS project. 

\section*{Acknowledgments}

The authors would like to thank Jennifer Mosher for the help fitting supernovae. 
This work had the financial support from CAPES and FAPESP Brazilian 
funding agencies. Funding for the SDSS and SDSS-II has been provided by the
Alfred P. Sloan Foundation, the Participating Institutions, the National
Science Foundation, the U.S. Department of Energy, the National Aeronautics
and Space Administration, the Japanese Monbukagakusho, the Max Planck Society,
and the Higher Education Funding Council for England. The SDSS Web Site is
\verb9http://www.sdss.org/9.

The SDSS is managed by the Astrophysical Research Consortium for the
Participating Institutions. The Participating Institutions are the American
Museum of Natural History, Astrophysical Institute Potsdam, University of
Basel, Cambridge University, Case Western Reserve University, University of
Chicago, Drexel University, Fermilab, the Institute for Advanced Study, the
Japan Participation Group, Johns Hopkins University, the Joint Institute for
Nuclear Astrophysics, the Kavli Institute for Particle Astrophysics and
Cosmology, the Korean Scientist Group, the Chinese Academy of Sciences
(LAMOST), Los Alamos National Laboratory, the Max-Planck-Institute for
Astronomy (MPIA), the Max-Planck-Institute for Astrophysics (MPA), New Mexico
State University, Ohio State University, University of Pittsburgh, University
of Portsmouth, Princeton University, the United States Naval Observatory, and
the University of Washington.

Funding for SDSS-III has been provided by the Alfred P. Sloan Foundation, 
the Participating Institutions, the National Science Foundation, and the U.S. 
Department of Energy Office of Science. The SDSS-III web site is http://www.sdss3.org/.

SDSS-III is managed by the Astrophysical Research Consortium for the Participating 
Institutions of the SDSS-III Collaboration including the University of Arizona, 
the Brazilian Participation Group, Brookhaven National Laboratory, University of 
Cambridge, Carnegie Mellon University, University of Florida, the French Participation 
Group, the German Participation Group, Harvard University, the Instituto de Astrofisica 
de Canarias, the Michigan State/Notre Dame/JINA Participation Group, Johns Hopkins 
University, Lawrence Berkeley National Laboratory, Max Planck Institute for Astrophysics, 
Max Planck Institute for Extraterrestrial Physics, New Mexico State University, New York 
University, Ohio State University, Pennsylvania State University, University of Portsmouth, 
Princeton University, the Spanish Participation Group, University of Tokyo, University of 
Utah, Vanderbilt University, University of Virginia, University of Washington, and 
Yale University. 

\bibliographystyle{mn2e}
\bibliography{main}

\appendix

\section{Determining sample requirements}
\label{sec:Contamination-estimation}

\subsection{Derivation of a guiding formula}

When deciding the requirements SNe Ia must fulfill in order to be considered
members of galaxy clusters, one must face a trade-off between completeness
and purity. Highly restrictive requirements will result in a small
but pure sample, while low requirements will result in a large but
contaminated sample. In order to guide our decision on the best type 
of sample to address our problem, suppose that some SNIa property $X$
has the average value $\bar{X}_{S}$ for SNe Ia in one sample $S$ and
$\bar{X}_{S'}$ in a different sample $S'$. One way to
determine if $S$ and $S'$ have distinct properties is by applying
a difference test such as the Z-test \citep{SprintHall90mn}: 

\begin{equation}
Z=\frac{\bar{X}_{S}-\bar{X}_{S'}}{\sqrt{\sigma_{\bar{X}_{S}}^{2}+\sigma_{\bar{X}_{S'}}^{2}}}\;,
\label{eq:ztestsimple}
\end{equation}
where $\sigma_{\bar{X}_{S}}^{2}$ and $\sigma_{\bar{X}_{S'}}^{2}$
are the variances of the mean of $X$ for the $S$ and $S'$ sample.
High $Z$ values indicate that the samples in question represent different
populations. 

Now consider that sample $S$ ($S'$) contains $N_{S}=N_{\mathrm{C}}+N_{\mathrm{F}}$
($N_{S'}=N_{\mathrm{C}}'+N_{\mathrm{F}}'$) SNe Ia, where $N_{\mathrm{C}}$ ($N_{\mathrm{C}}'$)
supernovae are actually in clusters and $N_{\mathrm{F}}$ ($N_{\mathrm{F}}'$)
are in field galaxies. Assuming that the variances for the cluster
and the field population are the same 
($\sigma_{X_{\mathrm{C}}}^{2}=\sigma_{X_{\mathrm{F}}}^{2}=\sigma_{X}^{2}$), that

\begin{equation}
\sigma_{\bar{X}_{S}}^{2} = \frac{\sigma_{X_{S}}^{2}}{N_{S}} = 
\frac{\sum_{i=1}^{N_{S}}(X_i-\bar{X}_{S})^2}{N_{S}(N_{S}-1)} \simeq \frac{\bar{X^{2}}_{S}-\bar{X}_{S}^2}{N_{S}}\;,
\label{eq:variance-approx}
\end{equation}
and that

\begin{equation}
\begin{aligned}
\bar{X}_{S}\simeq & \frac{N_{\mathrm{C}}\bar{X}_{\mathrm{C}}+N_{\mathrm{F}}\bar{X}_{\mathrm{F}}}{N_{S}}\;,\\
\bar{X}_{S'}\simeq & \frac{N_{\mathrm{C}}'\bar{X}_{\mathrm{C}}+N_{\mathrm{F}}'\bar{X}_{\mathrm{F}}}{N_{S'}}\;,\\
\bar{X^{2}}_{S}\simeq & \frac{N_{\mathrm{C}}\bar{X^{2}}_{\mathrm{C}}+N_{\mathrm{F}}\bar{X^{2}}_{\mathrm{F}}}{N_{S}}\;,\\
\bar{X^{2}}_{S'}\simeq & \frac{N_{\mathrm{C}}'\bar{X^{2}}_{\mathrm{C}}+N_{\mathrm{F}}'\bar{X^{2}}_{\mathrm{F}}}{N_{S'}}\;,
\end{aligned}
\label{eq:aprox-on-avg}
\end{equation}
where $\bar{X}_{\mathrm{C}}$, $\bar{X^{2}}_{\mathrm{C}}$, $\bar{X}_{\mathrm{F}}$ and $\bar{X^{2}}_{\mathrm{F}}$ are 
the true $X$ and $X^2$ average values for the cluster and field populations, 
we arrive at the formula:

\begin{equation}
Z\simeq\frac{(1-\phi_{S}-\phi_{S'})}{\sqrt{\frac{1}{N_{S}}+
\frac{1}{N_{S'}}+\delta^{2}\left[\frac{\phi_{S}(1-\phi_{S})}{N_{S}}+
\frac{\phi_{S'}(1-\phi_{S'})}{N_{S'}}\right]}}\delta\;,
\label{eq:Ztest-final}
\end{equation}
\begin{equation}
\delta\equiv\frac{\bar{X}_{\mathrm{C}}-\bar{X}_{\mathrm{F}}}{\sigma_{X}}\;,
\label{eq:delta-def}
\end{equation}
where $\phi_{S}\equiv\frac{N_{\mathrm{F}}}{N_{S}}$ and $\phi_{S'}\equiv\frac{N_{\mathrm{C}}'}{N_{S'}}$.
If we try to construct $S$ ($S'$) mainly from cluster (field) SNe Ia,
$\phi_{S}$ ($\phi_{S'}$) is interpreted as a contamination
fraction by field (cluster) supernovae. In our case, our tentative
cluster SNIa sample $S$ will dominate both contamination and Poisson
noise since $N_{S}\ll N_{S'}$ and cluster SNIa is a rarer event
than field SNIa. We then approximate Eq. \ref{eq:Ztest-final} to:

\begin{equation}
Z\simeq\frac{(1-\phi_{S})\sqrt{N_{S}}}{\sqrt{1+\phi_{S}(1-\phi_{S})\delta^2}}\delta\;.
\label{eq:Ztest-aprox}
\end{equation}
For a given $\delta$, we must find the sample $S$ membership requirements
(in this work, the maximum angular separation $\theta_{\mathrm{max}}^{(k)}$,
maximum redshift difference $z_{\mathrm{d}}$ and minimum probability of compatible
redshift with a true cluster $P_{\mathrm{min}}$) which maximize $Z$. To accomplish this task, we
must estimate how the requirements influences the contamination $\phi_{S}$.

\subsection{Estimating sample contamination }

\subsubsection*{For clusters with spectroscopic redshifts}

The choice of $\theta_{\mathrm{max}}^{(k)}$ was based on previous studies 
\citep[see][]{Dilday10mn,Mannucci08mn} and set to a maximum projected physical
separation of 1.5 Mpc. In this work we did not assess how
the choice of $\theta_{\mathrm{max}}^{(k)}$ affected the contamination, and
assumed that SNe Ia within this angular separation are necessarily
projected onto the cluster. The maximum redshift difference $z_{\mathrm{d}}$
for clusters with \emph{spec-z} was set to $0.005$, corresponding to a 
1500 $\mathrm{km\;s^{-1}}$ maximum difference in cluster member 
velocities. For a Gaussian velocity dispersion with 
$\sigma_v = 500\mathrm{km\;s^{-1}}$, this corresponds to a maximum difference 
of $3\sigma_v$. 

The contamination of our cluster SNIa sample by field SNe Ia, $\phi_{S}$, was calculated as:
\begin{equation}
\phi_{S}=\sum_{n=1}^{N_{S}}\left[(1-q_n)+q_n(1-p_{n})+q_np_{n}\Pi_{n}\right]\;,
\label{eq:contam-specz}
\end{equation}
where $q_n$ is the probability given by Table \ref{tab:cluster-purity} that 
the cluster onto which the supernova $n$ is projected is real, 
$p_{n}$ is the probability that the supernovae $n$ belonging
the the sample $S$ has a compatible redshift with the cluster, obtained
by Eq. \ref{eq:compatzspec}. Every SNIa in this sample must have
$q_np_{n}>P_{\mathrm{min}}$ (besides a angular separation less than $\theta_{\mathrm{max}}^{(k)}$); 
thus $P_{\mathrm{min}}$ controls both $N_{S}$ and $\phi_{S}$.

The $q_np_{n}\Pi_{n}$ term accounts for the fact that redshift compatibility
between the supernova and the cluster does not necessarily mean that
they are bound, since the observed redshift difference may be due
to cosmic expansion and thus reflect a comoving separation of $\sim$20 Mpc.
Therefore, the quantity $\Pi_{n}$ is the probability that a field SNIa
is able to fulfill our cluster membership requirements, and was estimated based on reported
SNIa rates, cluster luminosities and galaxy luminosity function \citep{Dilday10mn,Blanton03mn}.
Given the higher rates per luminosity of clusters and its higher luminosity
when compared with the field galaxies inside the volume comprised
by $\theta_{\mathrm{max}}^{(k)}$ and $z_{\mathrm{d}}$, this term is negligible when
compared to the sum of $(1-q_n)$, which is the probability for the cluster to be a
projection of field galaxies, and $q_n(1-p_{n})$, which is the probability that 
the cluster is real but SNIa $n$ does not have a compatible redshift.

\subsubsection*{For clusters with photometric redshift}

The choice of $\theta_{\mathrm{max}}^{(k)}$ was the same as for clusters with
spectroscopic redshift, while $z_{\mathrm{d}}$ and $P_{\mathrm{min}}$, in this case,
were both determined by maximising $Z$ in Eq. \ref{eq:Ztest-aprox}.
The contamination $\phi_{s}$ was determined by using the sample of
SNe Ia in clusters with spectroscopic redshift as a fiducial catalogue
and counting how many previously rejected SNe Ia were admitted using
the photometric redshift. The resulting values for $z_{\mathrm{d}}$ and $P_{\mathrm{min}}$
were then used as requirements for the clusters without spectroscopic
redshift. 

\subsection{Results for sample requirements}

The maximisation of $Z$ in Eq. \ref{eq:Ztest-aprox} was done separately for 
the different types of cluster redshift, and in both cases it was performed using 
the values $\delta\simeq 0$ as well as $\delta=1$. In all cases the value of 
$P_{\mathrm{min}}=0.50$ would maximize $Z$. This value is reassuring since it includes 
in our sample SNe Ia with a marginally higher change of belonging to a cluster 
and excludes them otherwise. The contamination by field SNe Ia in the \emph{spec-z} cluster 
sub-sample was estimated as 18 per cent. 

For clusters with only a photometric redshift, the difference in $\delta$ also did not affect 
the choice of $z_{\mathrm{d}}$, which was set to 0.030, resulting in a contamination of 42 per cent. The 
combination of both sub-samples was found to increase $Z$ and the final contamination rate 
for the cluster sample was estimated at 29 per cent. 

% Full cluster sample data table 

\clearpage
\pagestyle{empty}
\onecolumn
\begin{landscape}
\setlength{\LTcapwidth}{\textwidth}
\begin{center}
\begin{longtable}{ccccccccccccccc}
\caption{
All SNe Ia in the cluster sample. The columns are, from left to right: the SNIa's candidate ID, 
its right ascension $\alpha_{\mathrm{sn}}$, declination $\delta_{\mathrm{sn}}$ (J2000), redshift 
$z_{\mathrm{sn}}$, typing kind; the identifier (the SDSS objID of its BCG) of the GMBCG cluster 
hosting the SNIa; the SNIa's SALT2 'stretch' parameter $x_1$, its SALT2 colour $c$, its Hubble 
Residual (HR) obtained with the full sample (cluster + field) nuisance parameters; the SDSS-DR8 
objID of its host galaxy; its host's mass-weighted average age (in Gyr), mass (in log of solar 
masses) and specific star formation rate (sSFR) (in log of $\mathrm{yr}^{-1}$).}
\label{tab:fulltable} \\

\hline
\multicolumn{1}{c}{CID} & 
\multicolumn{1}{c}{$\alpha_{\mathrm{sn}}$} &
\multicolumn{1}{c}{$\delta_{\mathrm{sn}}$} & 
\multicolumn{1}{c}{$z_{\mathrm{sn}}$} & 
\multicolumn{1}{c}{Tp$^1$} & 
\multicolumn{1}{c}{Cluster BCG ID} & 
\multicolumn{1}{c}{$x_{1}$} & 
\multicolumn{1}{c}{$c$} & 
\multicolumn{1}{c}{HR} & 
\multicolumn{1}{c}{Host ID$^2$} & 
\multicolumn{1}{c}{Age$^3$} & 
\multicolumn{1}{c}{Mass$^3$} & 
\multicolumn{1}{c}{sSFR$^3$}
\\ \hline 
\endfirsthead

\multicolumn{15}{c}%
{\tablename\ \thetable{} -- continued.} \\
\hline 
\multicolumn{1}{c}{CID} & 
\multicolumn{1}{c}{$\alpha_{\mathrm{sn}}$} &
\multicolumn{1}{c}{$\delta_{\mathrm{sn}}$} & 
\multicolumn{1}{c}{$z_{\mathrm{sn}}$} & 
\multicolumn{1}{c}{Tp$^1$} & 
\multicolumn{1}{c}{Cluster BCG ID} & 
\multicolumn{1}{c}{$x_{1}$} & 
\multicolumn{1}{c}{$c$} & 
\multicolumn{1}{c}{HR} & 
\multicolumn{1}{c}{Host ID$^2$} & 
\multicolumn{1}{c}{Age$^3$} & 
\multicolumn{1}{c}{Mass$^3$} & 
\multicolumn{1}{c}{sSFR$^3$}
\\ \hline 
\endhead

\endfoot

\hline
\multicolumn{15}{l}{
$^1$ S -- spectroscoscopic typing; P -- photometric typing with host's \emph{spec-z} from SDSS-II; B -- photometric typing with host's \emph{spec-z} from BOSS.}\\
\multicolumn{15}{l}{
$^2$ Rows with no value correspond to SNIa with no identifiable host.}\\
\multicolumn{15}{l}{$^3$ Rows with no value correspond to hosts that did not pass the chi-squared minimum probability cut.}
\endlastfoot

822 & 40.5608 & -0.8622 & 0.2376 & B & 587731511541563584 & -0.48(53) & -0.115(45) & 0.28(23) & 1237657584950379049 & $4.55^{+1.96}_{-1.77}$ & $10.01^{+0.15}_{-0.14}$ & $-10.30^{+0.24}_{-0.46}$ \\
1166 & 9.3556 & 0.9733 & 0.3821 & S & 588015510347186388 & 1.4(1.1) & 0.005(64) & -0.23(31) & 1237663716555293384 & $6.60^{+2.10}_{-2.40}$ & $11.166^{+0.073}_{-0.11}$ & $-16.1^{+3.8}_{-24.6}$ \\
2855 & 16.1753 & -0.3564 & 0.2451 & B & 588015508739522825 & 0.64(45) & -0.051(43) & 0.01(23) & 1237663783667630515 & $3.91^{+1.89}_{-1.31}$ & $9.71^{+0.11}_{-0.11}$ & $-10.20^{+0.16}_{-0.15}$ \\
5717 & 17.8959 & -0.0058 & 0.2517 & S & 587731512605409559 & 1.36(36) & -0.030(28) & 0.22(19) & 1237666339189555766 & $2.64^{+1.00}_{-0.54}$ & $9.592^{+0.098}_{-0.075}$ & $-10.03^{+0.05}_{-0.09}$ \\
6300 & 37.4116 & -0.5501 & 0.3581 & B & 587731512077058240 & -3.95(04) & -0.130(25) & -0.27(19) & 1237657070089732863 & $4.92^{+1.99}_{-2.51}$ & $11.017^{+0.142}_{-0.138}$ & $-11.51^{+0.40}_{-0.87}$ \\
6560 & -38.5534 & 0.8498 & 0.2733 & B & 587730848501727276 & -1.24(88) & -0.034(79) & 0.09(28) & 1237663458852078081 & $2.64^{+1.16}_{-0.52}$ & $9.737^{+0.098}_{-0.075}$ & $-10.06^{+0.10}_{-0.09}$ \\
6743 & -19.1357 & -0.3667 & 0.3621 & B & 587734304344244570 & -1.7(1.0) & -0.101(94) & 0.05(32) & 1237663478725214894 & $5.10^{+2.27}_{-1.76}$ & $10.81^{+0.17}_{-0.15}$ & $-12.8^{+2.1}_{-15.7}$ \\
7802 & -14.4239 & 0.7292 & 0.4076 & B & 587731187271336219 & -0.06(89) & -0.233(57) & 0.32(27) & 1237663462604997493 & $5.50^{+1.99}_{-2.11}$ & $10.93^{+0.21}_{-0.17}$ & $-12.2^{+1.7}_{-14.6}$ \\
8160 & 33.7651 & -1.1004 & 0.4082 & B & 587731511538614462 & -0.6(1.2) & 0.07(11) & -0.48(32) & 1237663782601556483 & $3.53^{+2.13}_{-1.15}$ & $10.98^{+0.10}_{-0.12}$ & $-12.5^{+1.7}_{-10.0}$ \\
9467 & -31.0486 & 1.1808 & 0.2203 & S & 587731187800932733 & -1.24(46) & -0.171(50) & 0.27(22) & 1237678595929407536 & - & - & - \\
11172 & -37.5873 & -0.2022 & 0.1362 & P & 587730846891573582 & -1.21(42) & 0.067(50) & 0.35(21) & 1237663543142646320 & $2.93^{+0.45}_{-0.00}$ & $10.174^{+0.026}_{-0.001}$ & $-10.079^{+0.000}_{-0.029}$ \\
12971 & 6.6485 & -0.3021 & 0.2352 & S & 588015508735459404 & 0.96(38) & -0.084(34) & -0.06(20) & 1237663783663436008 & - & - & - \\
13511 & 40.6124 & -0.7941 & 0.2374 & S & 587731511541563584 & -1.40(43) & -0.117(38) & 0.26(21) & 1237663783141441728 & $5.05^{+2.08}_{-1.51}$ & $11.241^{+0.164}_{-0.032}$ & $-11.58^{+0.34}_{-0.51}$ \\
13689 & 4.0161 & 0.8075 & 0.2518 & S & 588015510344892595 & 1.19(42) & -0.112(33) & 0.11(20) & 1237657191980728518 & - & - & - \\
13757 & -9.8769 & -1.1578 & 0.2890 & S & 588015507654377685 & 0.95(46) & -0.176(42) & 0.18(22) & - & - & - & - \\
13952 & 4.6345 & 0.7886 & 0.3294 & B & 587731187279659200 & 3.0(1.3) & 0.045(72) & 0.65(37) & 1237657191980990962 & - & - & - \\
14340 & -14.1734 & -0.8553 & 0.2774 & P & 587731185123983494 & -0.58(52) & 0.017(50) & -0.19(23) & 1237656906348888177 & $6.54^{+2.00}_{-3.00}$ & $11.535^{+0.065}_{-0.143}$ & $-16.1^{+3.5}_{-20.1}$ \\
14444 & -23.2984 & -0.8157 & 0.2459 & B & 587731185119986013 & -1.15(36) & -0.186(39) & 0.33(21) & 1237656567585768071 & $7.45^{+2.40}_{-2.60}$ & $10.955^{+0.080}_{-0.102}$ & $-16.4^{+3.5}_{-25.5}$ \\
14624 & -36.5587 & 0.3459 & 0.4349 & B & 587731186724831882 & 1.89(85) & 0.008(63) & -0.08(30) & 1237663543679976707 & - & - & - \\
14984 & -46.1664 & -0.0928 & 0.1840 & S & 587731173306007979 & 1.01(37) & 0.063(33) & 0.14(21) & 1237663543138911569 & $2.86^{+0.80}_{-0.35}$ & $10.359^{+0.053}_{-0.034}$ & $-10.074^{+0.043}_{-0.064}$ \\
15287 & -36.0403 & -1.0574 & 0.2377 & S & 587730845818421799 & 0.92(30) & -0.045(29) & 0.11(19) & 1237656567043327064 & $8.03^{+1.96}_{-2.10}$ & $10.527^{+0.073}_{-0.091}$ & $-16.4^{+4.0}_{-30.1}$ \\
15354 & 6.7737 & -0.1260 & 0.2221 & S & 587731186206900338 & -2.34(38) & 0.074(51) & -0.30(22) & 1237657190908166490 & $8.69^{+1.50}_{-2.00}$ & $10.809^{+0.056}_{-0.070}$ & $-18.3^{+4.5}_{-29.0}$ \\
15648 & -46.2816 & -0.1948 & 0.1750 & S & 587731173306007979 & -1.50(42) & 0.094(51) & -0.13(20) & 1237663543138845573 & $11.19^{+0.40}_{-0.60}$ & $11.290^{+0.003}_{-0.030}$ & $-18.7^{+1.7}_{-39.4}$ \\
15803 & 18.2708 & -0.6718 & 0.4293 & B & 588015508203634864 & -1.4(1.1) & -0.106(97) & -0.03(37) & 1237663783131676936 & - & - & - \\
15897 & 11.6815 & -1.0329 & 0.1749 & S & 588015507663814778 & -2.52(31) & -0.026(40) & -0.08(20) & 1237657189836587309 & $9.60^{+1.60}_{-2.00}$ & $10.824^{+0.074}_{-0.060}$ & $-17.9^{+4.3}_{-35.9}$ \\
16021 & 13.8437 & -0.3888 & 0.1246 & S & 588015508738539674 & -0.30(18) & -0.039(25) & -0.08(17) & 1237663783666581814 & - & - & - \\
16103 & -47.0249 & -1.0501 & 0.2024 & P & 587730845813638207 & -1.44(31) & -0.093(34) & 0.01(19) & 1237656567038543385 & $7.40^{+2.04}_{-2.10}$ & $10.430^{+0.074}_{-0.106}$ & $-13.3^{+2.0}_{-21.9}$ \\
16160 & 24.7646 & -0.9524 & 0.3204 & B & 587731511534682250 & -1.7(1.0) & 0.11(10) & -0.17(36) & 1237657069547356652 & $6.24^{+2.06}_{-2.50}$ & $11.077^{+0.076}_{-0.146}$ & $-13.1^{+1.5}_{-21.3}$ \\
16213 & 8.9710 & 0.2584 & 0.2500 & S & 588015509273313434 & 0.28(51) & -0.113(35) & 0.09(21) & - & - & - & - \\
16467 & -31.4143 & 0.1183 & 0.2212 & B & 587734304875676035 & -2.25(50) & 0.004(52) & -0.28(22) & 1237663479256711274 & - & - & - \\
16473 & -31.3879 & 0.5885 & 0.2175 & S & 587734305412612580 & -1.51(76) & -0.179(43) & -0.05(19) & 1237663479793583244 & - & - & - \\
16482 & -31.2905 & 0.9336 & 0.2108 & S & 587734305949483196 & -1.03(97) & -0.073(42) & 0.06(23) & 1237678617404113136 & - & - & - \\
17820 & 16.1397 & -0.4665 & 0.3407 & B & 587731512067752240 & 0.49(56) & -0.087(56) & -0.13(26) & 1237666338651898752 & $7.56^{+2.00}_{-2.13}$ & $10.453^{+0.080}_{-0.099}$ & $-17.3^{+4.0}_{-25.0}$ \\
18301 & -11.0333 & 0.8163 & 0.3316 & B & 588015510338273460 & 1.12(98) & 0.145(79) & -0.23(33) & 1237663462606504597 & $3.46^{+1.89}_{-1.15}$ & $11.244^{+0.066}_{-0.127}$ & $-10.19^{+0.16}_{-0.43}$ \\
18325 & 8.9058 & 0.3700 & 0.2587 & S & 587731186744688861 & 0.34(40) & -0.086(42) & 0.07(21) & 1237657191445954792 & - & - & - \\
18362 & 10.1366 & -0.1820 & 0.2363 & B & 588015508736901265 & -0.91(43) & 0.015(52) & -0.06(22) & 1237657190909673715 & $7.41^{+1.68}_{-2.16}$ & $10.636^{+0.064}_{-0.065}$ & $-11.24^{+0.33}_{-0.43}$ \\
18375 & 11.5165 & -0.0104 & 0.1105 & S & 588015509274427469 & 0.96(16) & 0.044(23) & -0.17(18) & 1237657190910263423 & $6.659^{+1.11}_{-1.96}$ & $10.468^{+0.046}_{-0.152}$ & $-10.528^{+0.062}_{-0.173}$ \\
18389 & 34.1498 & 0.2295 & 0.2449 & B & 587731513149358332 & 0.81(63) & 0.391(60) & -0.38(25) & 1237666408457371795 & $2.11^{+0.03}_{-0.00}$ & $10.860^{+0.009}_{-0.000}$ & $-9.990^{+0.006}_{-0.000}$ \\
18606 & -12.9881 & 0.4823 & 0.3385 & B & 588015509800550549 & 0.28(94) & 0.129(74) & -0.32(31) & 1237663277922058907 & $2.80^{+1.77}_{-1.01}$ & $10.057^{+0.111}_{-0.102}$ & $-10.06^{+0.13}_{-0.22}$ \\
18767 & 4.5346 & 0.8087 & 0.3302 & B & 587731187279659200 & -1.54(75) & -0.074(81) & -0.30(31) & 1237657191980925614 & - & - & - \\
19341 & 15.8603 & 0.3314 & 0.2368 & S & 587731513141362889 & -1.44(39) & -0.009(43) & -0.11(21) & 1237666339725508846 & $7.54^{+1.40}_{-2.60}$ & $11.002^{+0.055}_{-0.092}$ & $-16.4^{+3.6}_{-25.8}$ \\
19708 & 42.1729 & 0.6547 & 0.2374 & B & 587731513689768062 & -0.78(36) & -0.045(39) & -0.42(19) & 1237657587098583274 & $7.04^{+2.90}_{-2.10}$ & $10.898^{+0.070}_{-0.097}$ & $-16.4^{+4.0}_{-25.8}$ \\
20111 & -5.6056 & 0.2481 & 0.2442 & S & 587731186738266460 & 0.16(74) & -0.001(44) & 0.05(22) & 1237666408439940055 & $7.07^{+1.94}_{-1.20}$ & $10.844^{+0.060}_{-0.066}$ & $-11.97^{+0.53}_{-0.93}$ \\
20232 & 7.0833 & -0.0581 & 0.2177 & B & 587731186207031493 & -1.12(58) & -0.009(40) & -0.25(20) & 1237657190908297506 & $9.24^{+0.90}_{-2.00}$ & $11.107^{+0.039}_{-0.062}$ & $-13.82^{+2.11}_{-29.30}$ \\
20723 & 41.2188 & -0.1074 & 0.3773 & B & 587731512615633103 & -2.26(82) & -0.113(97) & -0.08(33) & 1237657070628306973 & $5.64^{+2.00}_{-2.40}$ & $11.131^{+0.095}_{-0.112}$ & $-16.2^{+4.3}_{-20.4}$ \\
20757 & 43.6217 & 0.2215 & 0.3655 & B & 587731513153487020 & 1.4(1.6) & 0.212(81) & -0.44(35) & 1237657586562302517 & - & - & - \\
20768 & 40.6259 & -0.9711 & 0.2383 & S & 587731511541563584 & -1.2(1.1) & 0.046(63) & 0.24(26) & 1237657584950379205 & - & - & - \\
20882 & 16.9923 & 0.4637 & 0.3144 & B & 588015509813657766 & 0.59(94) & -0.102(56) & 0.29(26) & 1237663784741700254 & $4.32^{+2.11}_{-1.13}$ & $10.47^{+0.12}_{-0.10}$ & $-13.2^{+1.7}_{-12.8}$ \\
\end{longtable}
\end{center}
\end{landscape}
 
\end{document}